\documentclass[onecollarge]{svjour2}     % onecolumn "king-size"

\smartqed  % flush right qed marks, e.g. at end of proof

\usepackage{latexsym}
\usepackage{mathptmx}      % use Times fonts if available on your TeX system
\usepackage{graphicx}
\usepackage{amsmath}
\usepackage{verbatim}
\usepackage{amssymb}

\newcommand{\bra}[1]{\langle #1|}
\newcommand{\ket}[1]{|#1\rangle}

\journalname{Foundations of Physics}
\begin{document}

\title{Finite-time destruction of entanglement and non-locality by environmental influences}
%\subtitle{Do you have a subtitle?		\\ If so, write it here}
%\titlerunning{Short form of title}    % If too long for running head

\author{Kevin Ann \and Gregg Jaeger}


\institute{Kevin Ann \at
              Department of Physics, Boston University, Boston, MA 02215 \\
              \email{kevinann@bu.edu}           		%  \\
           \and
           Gregg Jaeger \at
             	Quantum Imaging Lab, Department of Electrical and Computer Engineering, 
             	and Division of Natural Sciences, Boston University, Boston, MA 02215 \\
              \email{jaeger@bu.edu}}

\date{Received: date / Accepted: date}

\maketitle

\begin{abstract}
Entanglement and non-locality are non-classical global characteristics of quantum states important
to the foundations of quantum mechanics.  Recent investigations have shown that environmental noise, 
even when it is entirely local in influence, can destroy both of these properties in finite time despite giving 
rise to full quantum state decoherence only in the infinite time limit. These investigations, which have 
been carried out in a range of theoretical and experimental situations, are reviewed here.


% MUST UPDATE!
% [PACS 2008 	- http://www.aip.org/pacs/pacs08/ASCII2008FullPacs.txt]
% (03.65.Ta 	- Foundations of quantum mechanics; measurement theory)
% (03.65.Ud 	- Entanglement and quantum non-locality)
% (03.67.-a 	- Quantum information)

\end{abstract}

%%%%%%%%%%%%%%%%%%%%%%%%%%%%%%%%%%%%%%%%%%%%%%%%%%%%%%%%%%%%%%%%%%%%%%%%%%%%%%%%%%%%%%%%%%%%%%%%%%%
%%%%%%%%%%%%%%%%%%%%%%%%%%%%%%%%%%%%%%%%%%%%%%%%%%%%%%%%%%%%%%%%%%%%%%%%%%%%%%%%%%%%%%%%%%%%%%%%%%%
%%%%%%%%%%%%%%%%%%%%%%%%%%%%%%%%%%%%%%%%%%%%%%%%%%%%%%%%%%%%%%%%%%%%%%%%%%%%%%%%%%%%%%%%%%%%%%%%%%%
\section{Introduction}\label{Introduction}

Entanglement and non-locality, which long have been of interest in the foundations of quantum theory, 
have been of renewed interest during the last two decades as the field of quantum information science 
emerged and matured. The techniques most often used in the latter context have recently been applied 
in the investigation of these properties, often with an eye toward the foundations of quantum theory. The 
investigation of the effects of noise on entanglement and non-locality is a potentially important example 
of this, because it could prove more significant for the foundations of quantum theory than the study of 
the simpler related effect of quantum decoherence, from which they have been shown during its course
to have important differences.

Entanglement and non-locality are distinct properties that are often conflated. Therefore, it is important 
to clarify their relationship as well as their relationships to quantum coherence.  Quantum coherence may be
destroyed  abruptly, for example, by strong projective measurements or asymptotically in time, for example, due 
to weak noise influences as a quantum system interacts with its environment.  It has been shown that 
even in the latter case, in which coherence disappears only in the infinite-time limit, entanglement and 
non-locality may still abruptly and completely disappear in finite time, phenomena named Entanglement 
Sudden Death (ESD) and Bell non-locality Sudden Death (BNSD), respectively.

After their relatively recent discovery, ESD and BNSD have been explored in a variety of 
contexts, theoretically and experimentally, in continuous and discrete systems, and when systems
are subject to noise, both classical and quantum in origin, which may affect both amplitude and phase. 
The physical systems that have been examined are diverse and include electrons on a solid state lattice 
subject to electromagnetic fields, photons traveling among collections of beamsplitters and mirrors, and 
atoms confined in electromagnetic cavities.  The results suggest that ESD and BNSD are 
ubiquitous phenomena which must be kept in mind in any realistic physical analysis.

This article provides a comprehensive survey of the investigation of ESD and BNSD.  
It is structured as follows. In Sec. \ref{Coherence}, we briefly outline the relevant details of the 
theory of quantum coherence and open quantum systems in order to properly contextualize ESD 
and BNSD. 
In Sec. \ref{Entanglement Sudden Death}, we survey work concerning ESD in both 
theoretical and experimental contexts.
In Sec. \ref{non-locality Sudden Death}, we turn attention to the investigation of BNSD and highlight 
how similar and distinct from ESD it is.
In Sec. \ref{Implications and summary}, we conclude with an examination of how ESD and BNSD may 
be of important to the interpretations and foundations of quantum 
mechanics and quantum information science.

%%%%%%%%%%%%%%%%%%%%%%%%%%%%%%%%%%%%%%%%%%%%%%%%%%%%%%%%%%%%%%%%%%%%%%%%%%%%%%%%%%%%%%%%%%%%%%%%%%%
%%%%%%%%%%%%%%%%%%%%%%%%%%%%%%%%%%%%%%%%%%%%%%%%%%%%%%%%%%%%%%%%%%%%%%%%%%%%%%%%%%%%%%%%%%%%%%%%%%%
%%%%%%%%%%%%%%%%%%%%%%%%%%%%%%%%%%%%%%%%%%%%%%%%%%%%%%%%%%%%%%%%%%%%%%%%%%%%%%%%%%%%%%%%%%%%%%%%%%%
\section{Coherence}\label{Coherence}

In recent decades, quantum coherence has most often been examined from the 
perspective of its destruction.  Decoherence has been shown to result in the 
degradation of quantum properties, corresponding to a transition from the quantum domain
to the classical domain. In general, decoherence may be classified into two main categories.  
The study of the first, intrinsic decoherence, follows an approach that modifies the standard 
quantum dynamics, leading to self-induced decoherence (SID); a stochastic term is added to the 
Schr$\ddot{\rm o}$dinger equation to induce spontaneous state-function collapse.  
Most studies of SID examine the following dephasing master 
equation for a state $\rho$ and Hamiltonian H \cite{Gardiner91,WM96,ACF06}, 
\begin{equation}
\frac{\partial \rho}{\partial t} 
= \frac{[H,\rho]}{i\hbar} - \frac{\tau}{\hbar^{2}}[H,[H,\rho]] \ , 
\label{masterEquation}
\end{equation}
$\tau$ being a theory-dependent parameter.  In the energy basis, the solution is given by
\begin{equation}
\rho(t) = \sum_{n,n' = 0}^{\infty} = 
\rho_{n,n'}e^{-i(E_{n} - E_{n'})t/\hbar}\exp\left[- \frac{\gamma(t)(E_{n} - E_{n'})^{2}}{\hbar^{2}} \right]\ket{E_{n}}\bra{E_{n'}} \ , \\
\end{equation}
with $\rho_{n,n'=0} = \bra{E_{n}}\hat{\rho}(0)\ket{E_{n}'}$, $\dot{\gamma} = \tau$. The state purity is accordingly
\begin{equation}
\mathcal{P}(t) = {\rm tr} \rho^{2}=\sum_{n,n'=0}^{\infty}|\rho_{n,n'}|^{2}\exp\left[ - 
\frac{2\gamma(t)(E_{n} - E_{n}')^{2}}{\hbar^{2}}
\right] \ .
\end{equation}
Provided that the temporal parameter $\tau$ is a constant, the degradation of state purity, 
and so state coherence, occurs exponentially in time.

The study of extrinsic decoherence, by contrast, uses standard quantum dynamics and considers the 
environment surrounding the system of interest in interaction with it; the environment is typically modeled 
as a system of much larger dimensionality than that of the system of interest.
Numerous classical and quantum correlations arise between the system 
and environment by unitary evolution.  As a result, the reduced system state obtained by tracing over 
the environmental state generally exhibits decoherence and may also exhibit ESD or BNSD.

%%%%%%%%%%%%%%%%%%%%%%%%%%%%%%%%%%%%%%%%%%%%%%%%%%%%%%%%%%%%%%%%%%%%%%%%%%%%%%%%%%%%%%%%%%%%%%%%%%%
%%%%%%%%%%%%%%%%%%%%%%%%%%%%%%%%%%%%%%%%%%%%%%%%%%%%%%%%%%%%%%%%%%%%%%%%%%%%%%%%%%%%%%%%%%%%%%%%%%%
\subsection{Open quantum systems and noise}

The models discussed in this review, with a few exceptions, are primarily based on open quantum 
systems theory, which has as a core assumption that quantum systems are not viewed in isolation, 
but only with respect to and within some environment, that is, following the extrinsic decoherence approach.  
Not only is the environment's interaction with the system inevitable, it is not confined to a 
perturbative influence; the environment may strongly affect the system's time evolution 
and determine which quantities are observable.  Even if the exact evolution equations of the quantum 
system coupled to the environment were solved, their complexity would preclude detailed and meaningful 
analysis in all but the simplest systems.

Generally, open quantum systems theory considers a quantum system of interest, $\rho_{\rm sys}$, 
together with its environment, $\rho_{\rm env}$, in a joint state 
$\rho_{\rm tot} = \rho_{\rm sys} \otimes \rho_{\rm env}$, that evolves unitarily and independently 
according to their own internal Hamiltonians $H_{\rm sys}$ and $H_{\rm env}$, respectively, with 
an interaction Hamiltonian $H_{\rm int}$, of the form
\begin{equation}
H_{\rm tot} = H_{\rm sys} \otimes \mathbb{I} + \mathbb{I} \otimes H_{\rm env} + H_{\rm int} \ .
\end{equation}
The statistical density matrix of the joint system $\rho_{\rm st}(t)$ evolves unitarily, 
$\rho_{\rm st}(t) = U(t) \rho_{\rm tot} U^{\dagger}(0)$, with the unitary operator $U(t)$ given by 
the Schr$\ddot{\rm o}$dinger solution,
$U\left(t\right) = \exp\left[-i\int_{0}^{t}{dt' H\left(t'\right)}\right]$. During the interval
$[0,t)$, classical and quantum correlations build up between the system and the environment, with 
its much larger number degrees of freedom, and spread throughout the joint Hilbert space.  
The time-evolved density operator averaged over noise fields is
$\rho\left(t\right) = \left\langle \rho_{\rm st}(t) \right\rangle_{\rm noise}$.  When attention 
is confined to the system of interest by tracing out the environment after unitary evolution, 
$\rho_{\rm sys} = {\rm tr}_{\rm env}\rho_{\rm tot}$, the influence of the environment is seen 
as noise that causes dephasing and amplitude damping, asymptotically leading to 
complete decoherence.  

For simplicity and without loss of generality, this process may be modeled more concretely for purposes 
of calculation for specific physical situations in terms of the operator sum decomposition.  The 
time-evolved density matrix is given by the following completely positive and trace preserving (CPTP) map,
\begin{eqnarray}
\rho\left(t\right) = \mathcal{E}\left[\rho\left(0\right)\right] =
\sum_{\mu = 1}^{N}\overline{E}_{\mu}^{\dagger}\left(t\right)\rho\left(0\right)
\overline{E}_{\mu}\left(t\right) \ ,
\end{eqnarray}
where the operators in the decomposition $\overline{E}_{\mu}$ satisfy the respective positivity 
and trace preserving relations,
$
\sum_{\mu}\overline{E}_{\mu}^{\dagger}(t)\overline{E}_{\mu}(t) = \mathbb{I} \ \ {\rm and} \ \  
\sum_{\mu}\overline{E}_{\mu}(t)\overline{E}^{\dagger}_{\mu}(t) = \mathbb{I}
$, \cite{Stinespring55,Kraus83}.
Provided that the specific forms of the operators $\overline{E}_{\mu}$ satisfy the CPTP relations, 
a set of non-unique operators that model a variety of quantum noise effects, such as dephasing 
and amplitude damping, as well as more general quantum evolution, can be constructed.  The 
exact forms of the sets of operators $\left\{ \overline{E}_{\mu}(t) \right\}$ are provided here if
needed for clarity.  
%Many models for the investigation of entanglement loss assume open quantum 
%systems dynamics and completely positive maps, which provide physically motivated mathematical 
%constraints related to non-classical probability distributions.  
An informative overview of this open quantum systems approach has been given by Benatti 
and Floreanini \cite{BF05}.

Open quantum systems theory has been central to a better understanding of decoherence.  
This approach is motivated by the fact that environmental effects cannot always be treated as perturbative 
corrections.  Despite environmental noise causing a quantum system to decohere only 
asymptotically as $t \rightarrow \infty$, this noise may in finite time destroy uniquely 
non-classical properties such as entanglement or non-locality, that is, produce ESD or 
BNSD respectively.

%%%%%%%%%%%%%%%%%%%%%%%%%%%%%%%%%%%%%%%%%%%%%%%%%%%%%%%%%%%%%%%%%%%%%%%%%%%%%%%%%%%%%%%%%%%%%%%%%%%
%%%%%%%%%%%%%%%%%%%%%%%%%%%%%%%%%%%%%%%%%%%%%%%%%%%%%%%%%%%%%%%%%%%%%%%%%%%%%%%%%%%%%%%%%%%%%%%%%%%
\subsection{Coherence measures}

There exist a number of useful indicators of quantum coherence, both quantitative measures such as state 
purity and fidelity as well as qualitative methods such as the mere existence of off-diagonal 
terms in the density matrix. Those that are frequently used in the study of decoherence and disentanglement
are summarized in this section.

The purity of a state $\rho$ is an easily computable measure of quantum coherence, which is given by 
\begin{equation}
\mathcal{P}(\rho) = {\rm tr} \ \rho^{2} \ .
\end{equation}  
For a $d$-dimensional system, $1/d \leq\mathcal{P}(\rho) \leq 1$; the lower bound $1/d$ is reached only 
for the completely mixed state exhibiting no coherence, whereas the upper bound $1$ is saturated by 
any pure state. As a state decoheres, its purity is reduced. The greatest degree of of decoherence occurs
when a pure state, with initial unit purity 1 later reaches purity $1/d$.

The fidelity coherence measure for arbitrary mixed state density matrices $\rho_{\rm 1}$ and $\rho_{\rm 2}$, is given by
\begin{equation}
{\rm F}(\rho_{\rm 1},\rho_{\rm 2}) = \left[ {\rm tr} \left(   
\sqrt{\sqrt{\rho_{\rm 2}} \ \rho_{\rm 1}\sqrt{\rho_{\rm 2}}}
\right)\right]^{2} \ ,
\end{equation}
with $0 \leq {\rm F}(\rho_{\rm 1},\rho_{\rm 2}) \leq 1$.  The upper bound $1$ indicates that 
$\rho_{\rm 1}$ and $\rho_{\rm 2}$ are indistinguishable and the lower bound $0$ indicates that
$\rho_{\rm 1}$ and $\rho_{\rm 2}$ are orthogonal; the fidelity may be used in certain circumstances
to find the time of the loss of coherence of a state from to its initial state by taking
$\rho_{\rm 1} = \rho(t = 0)$ and $\rho_{\rm 2} = \rho(t)$ represent the initial state and 
the later state, respectively.

Although these measures are often suitable for quantifying the amount of quantum state 
coherence present, in many cases the concern is only whether quantum coherence exists or not;
for that purpose, it is sufficient to observe the off-diagonal elements, only 
when they are all zero is there no state coherence in the particular basis in which they are 
represented.

%%%%%%%%%%%%%%%%%%%%%%%%%%%%%%%%%%%%%%%%%%%%%%%%%%%%%%%%%%%%%%%%%%%%%%%%%%%%%%%%%%%%%%%%%%%%%%%%%%%
%%%%%%%%%%%%%%%%%%%%%%%%%%%%%%%%%%%%%%%%%%%%%%%%%%%%%%%%%%%%%%%%%%%%%%%%%%%%%%%%%%%%%%%%%%%%%%%%%%%
%%%%%%%%%%%%%%%%%%%%%%%%%%%%%%%%%%%%%%%%%%%%%%%%%%%%%%%%%%%%%%%%%%%%%%%%%%%%%%%%%%%%%%%%%%%%%%%%%%%
\section{Entanglement sudden death}\label{Entanglement Sudden Death}

Let us now consider work specifically focusing on Entanglement Sudden Death (ESD).
Sec. \ref{Preliminaries} briefly reviews required background, including the most often used entanglement 
measures and the classification of quantum states.  
Sec. \ref{First discoveries} discusses the initial discoveries of ESD in both continuous and discrete variable 
systems showing that decoherence and disentanglement are distinct phenomena in that they
proceed at qualitatively different rates.
Sec. \ref{Theoretical developments} surveys the subsequent theoretical development of the study
of ESD.
Sec. \ref{Breadth of ESD} discusses ESD as found in a range of different physical systems.
Finally, experimental confirmation of ESD and related experimental issues are described in 
\ref{ESD: Empirical evidence and experimental proposals}.

%%%%%%%%%%%%%%%%%%%%%%%%%%%%%%%%%%%%%%%%%%%%%%%%%%%%%%%%%%%%%%%%%%%%%%%%%%%%%%%%%%%%%%%%%%%%%%%%%%%
%%%%%%%%%%%%%%%%%%%%%%%%%%%%%%%%%%%%%%%%%%%%%%%%%%%%%%%%%%%%%%%%%%%%%%%%%%%%%%%%%%%%%%%%%%%%%%%%%%%
\subsection{Preliminaries}\label{Preliminaries}

In order to investigate Entanglement Sudden Death (ESD), one must bring to bear appropriate measures of 
entanglement, which are described in this section after introducing the relevant entanglement clases.

%%%%%%%%%%%%%%%%%%%%%%%%%%%%%%%%%%%%%%%%%%%%%%%%%%%%%%%%%%%%%%%%%%%%%%%%%%%%%%%%%%%%%%%%%%%%%%%%%%%
\subsubsection{Entanglement classes}\label{Entanglement classes}

Because entanglement is a global property of a state and non-increasing under local operations, 
there are natural state classifications provided by state behavior under local operations.  
Since classical communication also cannot affect entanglement, entangled state classification 
can likewise involve local operations in conjunction with classical communication.  A rigorous 
mathematical characterization of state classification was undertaken by D$\ddot{\rm u}$r, Vidal, 
and Cirac \cite{DVC00}.  Their development was based on states that could be obtained from each 
other through invertible local operators (ILOs), for example, in a three-qubit system,
\begin{equation}
\ket{\phi} = A \otimes B \otimes C \ket{\psi} 
\ \ \ \ \ {\rm and} \ \ \ \ \ \ket{\psi} = A^{-1} \otimes B^{-1} \otimes C^{-1}\ket{\phi} \ .
\end{equation}
If this relation holds for two  states, they are equivalent under 
local operations and classical communication (LOCC).  If this transformation can occur with 
any finite probability, no matter how small, then the two states 
have equal entanglement under ``stochastic'' local operations and classical communication (SLOCC).  

For two-qubit states, there exists only one equivalence class of interconvertible entangled pure states,
that of the Bell states.  For three-qubit systems, the SLOCC classification scheme identifies two distinct 
classes of genuinely tripartite entangled pure states.  The W class, which retains maximum bipartite entanglement 
after tracing out one of the qubits, is written generically as
\begin{equation}
\ket{W^{\rm g}} = \bar{a}_{1}\ket{001} + \bar{a}_{2}\ket{010} + \bar{a}_{4}\ket{100} \ .
\end{equation}
The GHZ class, represented by
\begin{equation}
\ket{GHZ^{\rm g}} = \bar{a}_{0}\ket{000} + \bar{a}_{7}\ket{111} \ ,
\end{equation}
is genuinely tripartite entangled and exhibits no entanglement at the bipartite level.  These 
genuinely tripartite entangled classes cannot be converted into one other, even with a finite 
probability, and are obtainable by projections from the generic tripartite state 
\begin{equation}
\ket{\rm \Psi_3} = \bar{a}_{0}\ket{000} + \bar{a}_{4}\ket{100} + 
\bar{a}_{5}\ket{101} + \bar{a}_{6}\ket{110} + \bar{a}_{7}\ket{111} \label{Psi3}
\end{equation} 
in $\mathcal{H}_{\rm ABC} = \mathcal{H}_{\rm A}\otimes\mathcal{H}_{\rm B}\otimes\mathcal{H}_{\rm C}$, where the complex coefficients
satisfy $\sum_{i}|\bar{a}_{i}|^2 = 1$.

Although many classifications exist for the multipartite system, studies in ESD and BNSD have  
primarily focused on the generalized $n$-qubit W class and GHZ class of states.  The generalized 
W-class is represented by 
\begin{equation}
\ket{W_{N}} = \frac{1}{\sqrt{n}}(\ket{100\ldots00} + \ket{010\ldots00} + \ldots + \ket{000\dots01}) \ ,
\end{equation}
and is characterized by its retaining maximal entanglement under the tracing out of a single qubit 
as well as exhibiting genuine $n$-partite entanglement.  The generalized GHZ-class is represented by 
\begin{equation}
\ket{GHZ_{N}} = \frac{1}{\sqrt{2}}(\ket{000\ldots00} + \ket{111\ldots11}) \ , 
\end{equation}
and is, in contrast, characterized by its genuinely $n$-partite entanglement, but exhibiting no 
entanglement when any qubit is traced over.

The classification of pure three-qubit states has been extended to the classification of mixed 
three-qubit states \cite{ADLS01}.  The generic class $\ket{\Psi_{3}}$ in Eq. \ref{Psi3} may 
be transformed with a finite probability through local operations into the GHZ class of 
states $\ket{GHZ^{\rm g}}$.  The GHZ states may themselves be transformed into the other 
tripartite classes: GHZ, W, biseparable (B), and separable (S) states through stochastic 
positive-operator-valued measures (POVMs) in the following hierarchy \cite{DVC00},
${\rm S} \subset {\rm B} \subset {\rm W} \subset {\rm GHZ}$.  Although there exist some classes 
of multipartite entangled mixed states, such as a generalized $n$-qubit Werner state composed of a 
maximally mixed state and an $n$-qubit GHZ-state, these have yet to be considered in the discussion 
of ESD from noise acting on a generalized multipartite state.  

All these entangled-state classifications are based on the requirement that local operations do not 
increase the entanglement of a state because it is global in character. This makes the discovery of ESD 
and BNSD under local noise, although clearly not in conflict with this requirement, somewhat remarkable
because a global property is seen to be destroyed by local influences alone.  Indeed, even very weak 
local noises are capable of destroying these local properties when acting on a range of initially entangled states.

%%%%%%%%%%%%%%%%%%%%%%%%%%%%%%%%%%%%%%%%%%%%%%%%%%%%%%%%%%%%%%%%%%%%%%%%%%%%%%%%%%%%%%%%%%%%%%%%%%%
\subsubsection{Entanglement measures}\label{Entanglement measures}

For studies involving two-qubit entanglement, concurrence $C(\rho)$ and the closely related 
entanglement of formation $E_{f}(\rho)$ are the measures most often used, because they are valid 
for both pure and mixed states.  For a two-qubit density matrix $\rho_{AB}$, the concurrence is  
\begin{equation}
C\left(\rho_{\rm AB}\right) = \max \left[0, \sqrt{\lambda_{1}} -
\sqrt{\lambda_{2}} - \sqrt{\lambda_{3}} - \sqrt{\lambda_{4}}\right] \ ,
\end{equation}
where the argument of the concurrence function,
$\Lambda \equiv \sqrt{\lambda_{1}} - \sqrt{\lambda_{2}} - \sqrt{\lambda_{3}} - \sqrt{\lambda_{4}}$, 
is a function of the eigenvalues $\lambda_{i}$ ($i = 1,2,3,4$), ordered by decreasing 
magnitude, of the matrix, 
$
\tilde{\rho}_{\rm AB} = \rho_{\rm AB}\left(\sigma_{y}^{{\rm A}}\otimes\sigma_{y}^{{\rm B}}\right) 
\rho_{\rm AB}^{\ast} \left(\sigma_{y}^{{\rm A}}\otimes\sigma_{y}^{{\rm B}} \right)
$, 
where $\rho_{\rm AB}^{\ast}$ is the complex conjugate of $\rho_{\rm AB}$, $\sigma_{y}^{{\rm A}({\rm B})}$ 
the standard Pauli matrix acting on qubit A(B) \cite{HW97,Wootters98}, and noting $C_{\rm AB} = C(\rho_{\rm AB})$ 
to simplify notation for later use;  the canonical measure of entanglement, the entanglement of formation, 
can be written in terms of the concurrence as follows.
\begin{equation}
E_{f}\left(\rho_{\rm AB}\right) = h\left[ \left(1 + \sqrt{1 - C_{\rm AB}^{2}}  \right)\Big/ 2 \right],
\end{equation}
where $h\left(x\right) = -x\log_{2}{x} - \left(1-x\right)\log_{2}{\left(1-x\right)}$.

The entanglement of both 2$\times$2 and 2$\times$3 complex-dimensional systems is based on the 
Peres-Horode$\check{\rm c}$ki criterion: if the partial transposition of the
joint system density matrix with respect to a subsystem has one or more negative eigenvalues, 
it is entangled \cite{Peres96,Horodecki97}.  This qualitative criterion can be quantified via the negativity, 
$\mathcal{N}(\rho)$, given by
\begin{equation}
\mathcal{N}(\rho_{\rm AB}) = \sum_{k} |\lambda^{\rm T_A(-)}_{k}|\ , 
\end{equation}
where $\lambda^{\rm T_A(-)}_{k}$ are the negative eigenvalues of the partial transpose of $\rho_{\rm AB}$ 
with respect to subsystem A, taken to be the lower-dimensionality one in the 2$\times$3 case  \cite{VW02,HHH96}.
Alternatively, but also 
based on the Peres-Horode$\check{\rm c}$ki criterion, Huang and Zhu have derived necessary and sufficient 
conditions for ESD based on a principle minor method and the negativity of eigenvalues \cite{HZ07,HZ08}.

The existence of entanglement measures that are valid for both pure and mixed states is necessary
to proceed to larger numbers of dimensions or subsystems.  For systems beyond the two-qubit and
qubit-qutrit cases, such measures are known to exist so far only in special circumstances, for 
example, when extra symmetries are present.  
For this reason, studies of ESD are not easily extendible to larger Hilbert spaces.  Furthermore, 
entanglement must be understood as differing from non-locality.  Already in the two-qubit system, 
there exist situations in which entanglement and non-locality differ,  
the Werner states can exhibit entanglement without violating a Bell inequality.
The closest larger case for a bipartite system is the symmetric 3x3-dimensional case.  Although 
there is no generalized entanglement measure known to exist so far for arbitrary mixed-state 
two-qutrit entanglement, 
Caves and Milburn \cite{CM00} derived the separability condition for a two-qutrit Werner-like 
state, $\rho_{\epsilon}$, composed of the maximally mixed component $\mathbb{I}_{9}/9$, 
and a maximally entangled component $\ket{\Psi} = (\ket{11} + \ket{22} + \ket{33})/\sqrt{3}$,
\begin{equation}
\rho_{\epsilon} = \frac{(1 - \epsilon)}{9}\mathbb{I}_{9} + \epsilon \ket{\Psi}\bra{\Psi} \ ,
\end{equation}
with $0 \leq \epsilon \leq 1$.  This state is separable, that is, not entangled, if and only if 
$\epsilon \leq 1/4$.

Although a mixed-state entanglement measure exists for the next closest symmetric dimensional 
increment beyond the qubit-qutrit case, a general entanglement measure has not yet so far been 
shown to exist for bipartite states of yet larger dimensions for arbitrary mixed states. However, a measure 
derived by Terhal and Vollbrecht \cite{TV00}, which included an ansatz later rigorously proven 
in \cite{FJ06}, does exist for the special case of $d \times d$ isotropic states, those invariant 
under transformations of the form $U \otimes U^{\ast}$, where $U$ is unitary \cite{HH99}.
The $d \times d$-dimensional isotropic states $\rho_{\rm iso}(d)$ consist of a convex combination 
of a maximally mixed density $(d^{-2})\mathbb{I}_{d^{2}}$ and a maximally entangled density
$P(\ket{\Psi(d)})\equiv\ket{\Psi(d)}\bra{\Psi(d)}$, that can be written as
\begin{equation}
\rho_{\rm iso}(d) = 
\left(\ \frac{1 - F}{d^{2} - 1} \right) \mathbb{I}_{d^{2}} + \left(\ \frac{F d^{2} - 1}{d^{2} - 1} \right) P(\ket{\Psi(d)}) \ ,
\end{equation} 
where $d>2$, $\mathbb{I}_{d^{2}}$ is the $d^{2} \times d^{2}$ identity matrix, 
$\ket{\Psi(d)} = (1/\sqrt{d})\sum_{i =1}^{d}{\ket{i} \ket{i}}$; fidelity
$
F(\rho_{\rm iso}(d), P(\ket{\Psi(d)})) = {\rm tr}\left(\rho_{\rm iso}(d) P(\ket{\Psi(d)})\right)
$ compare with \cite{Jozsa94}, 
with $0 \leq F(\rho_{\rm iso}(d), P(\ket{\Psi(d)})) \leq 1$ in the definition of isotropic
states \cite{HH99}.  $\rho_{\rm iso}(d)$ is separable if and only if 
$F\left(\rho_{\rm iso}(d),P(\ket{\Psi})\right)\leq F_{\rm critical}(d)\equiv d^{-1}$.  
The entanglement of formation for the isotropic states $\rho_{\rm iso}(d)$ for $d>2$,
\begin{equation}
E_{\rm f}\left(\rho_{\rm iso}\right) =
\begin{cases} 
0 																										& F \leq \frac{1}{d}\ , \\
R_{1, d-1}\left( F \right),       										& F \in \left[\frac{1}{d}, \frac{4(d-1)}{d^{2}} \right]\ , \\
\frac{d \log(d-1)}{d-2}\left( F - 1 \right) + \log d, & F \in \left[\frac{4(d-1)}{d^{2}}, 1\right]\ ,
\end{cases}
\end{equation}
where $R_{1, d-1}\left( F \right) = H_{2}\left(\xi(F)\right) +
\left[1 - \xi(F) \right]\log_{2}(d-1)$, \ $H_{2}(x) = -x\log_{2} (x)
- (1-x) \log_{2} (1- x)$, \ and $\xi(F) = \frac{1}{d} \left[
\sqrt{F} + \sqrt{(d-1)(1-F)}\right]^{2}$, {\it cf.} \cite{TV00,FJ06}.  

Other more general quantities relating to entanglement exist for large finite dimensions 
bipartite states, such as the generalized concurrence and entanglement of formation derived 
in \cite{CAF05a,CAF05b}.  These measures give a lower bound for entanglement content, 
but are often difficult to make use of due to the amount of computation involved in performing
the minimization and maximization required.  For increasing dimensions of the 
discrete-variable $d \times d$ system, one approaches bipartite 
continuous-variable systems of infinite dimensions, which are well-described by the 
Wigner function in terms of momentum $p_{i}$ and position $x_{i}$, 
\begin{equation}
W(p_{1},x_{1},p_{2},x_{2}) = \sum_{i}p_{i}W_{i}^{\rm A}(p_{1},x_{1})W_{i}^{\rm B}(p_{2},x_{2}) \ .
\end{equation}
In that limit, the continuous-variable analogue of the Peres-Horode$\check{\rm c}$ki 
condition is that, upon mapping $p_{2} \rightarrow -p_{2}$, the state remains entangled 
if the Wigner function remains a Wigner function.
This criterion was proven to be necessary and sufficient for a Gaussian bipartite system 
\cite{Simon00}.  Other necessary and sufficient conditions were derived for the Gaussian 
bipartite cases as well, such as those based on the variances of EPR-like operators 
\cite{DGCZ00}.  Some progress towards entanglement measures for multipartite continuous 
states has been made, but only for very limited cases that aren't readily applicable
for ESD studies.  For that reason, the discussion here of entanglement in 
continuous-variable systems is limited to the bipartite case.

Returning to the discrete case, for a three-qubit system, pure-state entanglement can be measured by the three-tangle 
$\tau_{\rm ABC}$ for subsystems A, B, and C, 
\begin{equation}
\tau_{\rm ABC} = C_{\rm A[BC]}^{2} - C_{\rm AB}^{2} - C_{\rm AC}^{2} \ , 
\end{equation}
where the bipartite concurrences involved are calculated in the usual way \cite{CKW00}. Although
this measure is not defined for mixed states, it is useful for the identification of {\em classes} of 
genuinely tripartite entangled of states, such as the GHZ and W class.  One form for a 
class of entanglement measures for mixed multipartite quantum states was developed in \cite{MKB05}, 
which is expressed in terms of all reduced density matrices $\rho_{k}$ from a state $\Psi$:
\begin{equation}
C_{N}(\Psi) = 2^{1 - N/2}\sqrt{(2^{N} - 2)\bra{\Psi}\ket{\Psi}^{2} - \sum_{k} \ {\rm tr}\ \rho_{k}^{2}}\label{nConcurrence} \ .
\end{equation}
The main difficulty with multipartite entanglement measures is the lack of easily parameterized 
closed-form expressions for arbitrary and mixed states. 
Furthermore, there exist partition-dependent qualitative types of entanglement, for example, different 
sets of $k$-partite divisions within a full $n$-partite system. Hence, one is led instead to considering 
multipartite non-classical correlations in terms of the non-local of correlations rather than entanglement,
as has been done to extend the sudden death of global state properties beyond ESD to BNSD, which
is discussed further below.

%%%%%%%%%%%%%%%%%%%%%%%%%%%%%%%%%%%%%%%%%%%%%%%%%%%%%%%%%%%%%%%%%%%%%%%%%%%%%%%%%%%%%%%%%%%%%%%%%%%
\subsection{First discoveries}\label{First discoveries}

Addressing the relationship between decoherence and disentanglement, that is, the nature of the loss of entanglement in relation to the loss of state coherence, was an important step toward the discovery of Entanglement Sudden Death (ESD) and Bell non-locality sudden death (BNSD). Intuitively, both of these 
quantum effects are expected to behave very similarly. ESD is the extreme case in which coherence persists 
asymptotically whereas the entanglement is entirely eliminated in finite time. In particular, the discovery that 
they may decay at different rates was a hint that ESD can occur.  

\begin{figure}[htbp]
	\centering
	\includegraphics[width=0.60\textwidth]{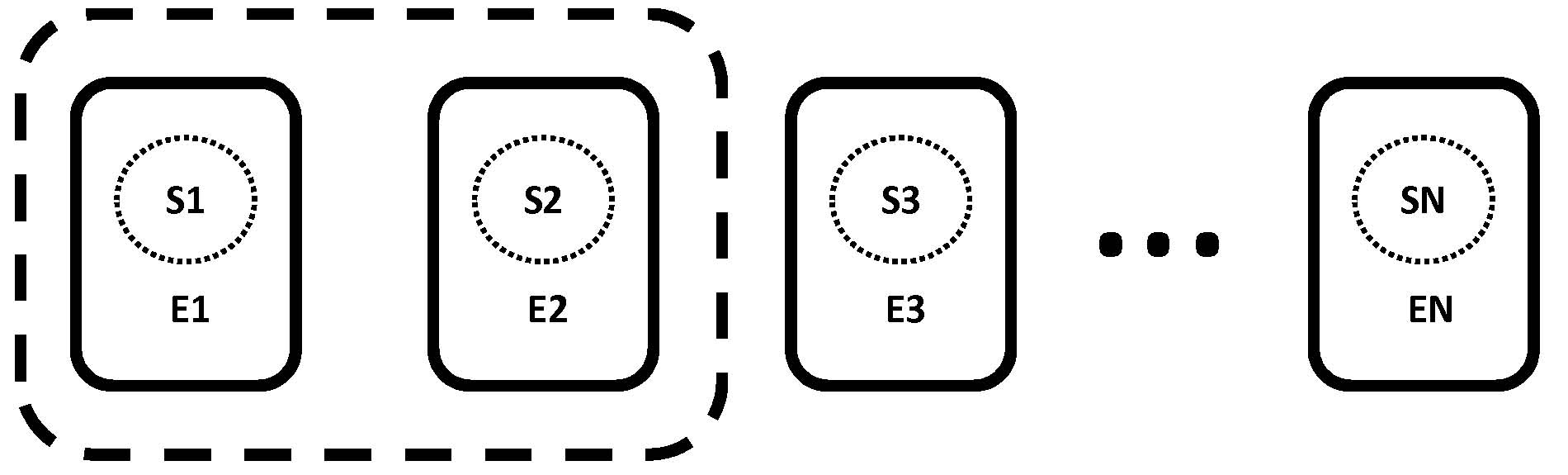}
	\caption{
	Correlated subsystems S$k$ each in their own noisy environments E$k$ ($k = 1, 2, ..., n$).  
	The region enclosed by the dashed line represents scenarios typical of two-qubit ESD studies and may 
	represent, for example, two electrons on separate positions of a solid state lattice each subject to 
	localized noise influences.  Various studies may stipulate further conditions, including: 
	(i) each localized subsystem-environment S$k$-E$k$ is noninteracting and nonlocally separated from the 
	others, with only information, entanglement, and nonlocal correlations connecting them, 
	(ii) subsystems may share a common environment that affects them collectively, 
	(iii) different sorts of noise that are present, such as quantum, classical, dephasing, or 
	amplitude damping, and
	(iv) existence and strength of couplings between subsystems and environments.}
	\label{fig01-locales}
\end{figure}

An important early step was taken by Yu and Eberly using an exactly solvable toy model consisting 
of a bipartite two-level system coupled to a purely phase reducing reservoir \cite{YE02}.  Their main discovery 
was that entanglement, which they measured using concurrence, decayed at a different rate from the coherence
as measured by the reduction of off-diagonal density matrix elements.  Specifically, the timescale of 
disentanglement was always less than or equal to the timescale of decoherence, $t_{\rm dis} \leq t_{\rm dec}$.
Other such studies followed to clarify the relationship between the rates of decoherence and disentanglement 
including \cite{YE03,TPA05,AJ07a,AJ07c}.  Fig. \ref{fig01-locales} shows the relationships 
between subsystems and their environments for a variety of studies.

The historically first discovery of ESD was that of Di$\acute{{\rm o}}$si in a bipartite two-level 
electron system subjected to a classical white dephasing environment of magnetic fields \cite{Diosi03}.  
This model includes depolarizing noise with decoherence timescale $\tau$, the Pauli spin vector 
$\stackrel{\rightharpoonup}{\sigma}$ of $2 \times 2$ Pauli spin matrices, and system density $\rho$,
\begin{equation}
\frac{d\rho}{dt} = 
- \frac{1}{4\tau}\left[\stackrel{\rightharpoonup}{\sigma},
\left[\stackrel{\rightharpoonup}{\sigma},\rho \right]\right] \ , 
\end{equation}
which can be interpreted as the statistical average of the wavefunction obeying the 
Schr$\ddot{\rm o}$dinger equation,
\begin{equation}
\frac{d\psi}{dt} = 
- \frac{i}{2}\stackrel{\rightharpoonup}{\omega}\stackrel{\rightharpoonup}{\sigma}\psi, 
\end{equation}
with $\stackrel{\rightharpoonup}{\omega}$ representing the random magnetic field producing
the noise. Using theorems in \cite{HSR03,Ruskai03} concerning disentangling quantum channels, 
complete disentanglement was shown by Di$\acute{{\rm o}}$si to occur upon satisfaction of 
some conditions proved true using the following arguments.  A nonseparable, that is, entangled 
state is defined as one that cannot be written as
\begin{equation}
\rho = \sum_{j} p_{j} \rho_{{\rm sys},j} \otimes \rho_{{\rm env},j} \ , 
\end{equation}
where $\left\{ p_{j} \right\}$, $\left\{ \rho_{{\rm sys},j} \right\}$, $\left\{ \rho_{{\rm env},j} \right\}$
are the set of weights, system states, and environmental states, respectively.  A linear map 
$\mathcal{M}(t)$ that time-evolves an initial state in time, $\rho(t) = \mathcal{M}(t)\rho(0)$, 
is called entanglement breaking if and only if
\begin{eqnarray}
\mathcal{M}\rho_{0} = \sum_{j}p_{j}\rho_{j} \ , \ \ \ \ \ p_{j} = {\rm tr}[P_{j}\rho] \ ,
\end{eqnarray}
where the set $\left\{ \rho_{j}\right\}$ are densities independent of the original state $\rho_{0}$,
and the probabilities $\left\{ p_{j} \right\}$ are obtained from $\left\{ P_{j} \right\}$, the 
set of positive operator valued measures (POVMs) satisfying 
$P_{j} \geq 0$ and $\sum_{j}P_{j} = \mathbb{I}$,
where $\mathbb{I}$ is the identity matrix.  The solution to this evolution equation can be written as
\begin{equation}
\rho(t) \equiv \mathcal{M}(t)\rho(0) = \frac{1}{2}\left[\mathbb{I}  + e^{-t/\tau} 
\stackrel{\rightharpoonup}{\sigma} \ {\rm tr}\left[ \stackrel{\rightharpoonup}{\sigma}\rho(0)\right] \right] \ , 
\end{equation}
where $\tau$ is the phase decay timescale and $\stackrel{\rightharpoonup}{\sigma}$ is the 
Pauli vector. ESD takes place if and only if $3e^{-t/\tau} \leq 1$.  Di$\acute{{\rm o}}$si proved 
this to be true by a constructive proof \cite{Diosi03} demonstrating the existence of ESD at a 
disentanglement time
\begin{equation}
t_{\rm dis} \geq \tau \ln 3 \ .
\end{equation}

\begin{figure}[htbp]
	\centering
	\includegraphics[width=0.60\textwidth]{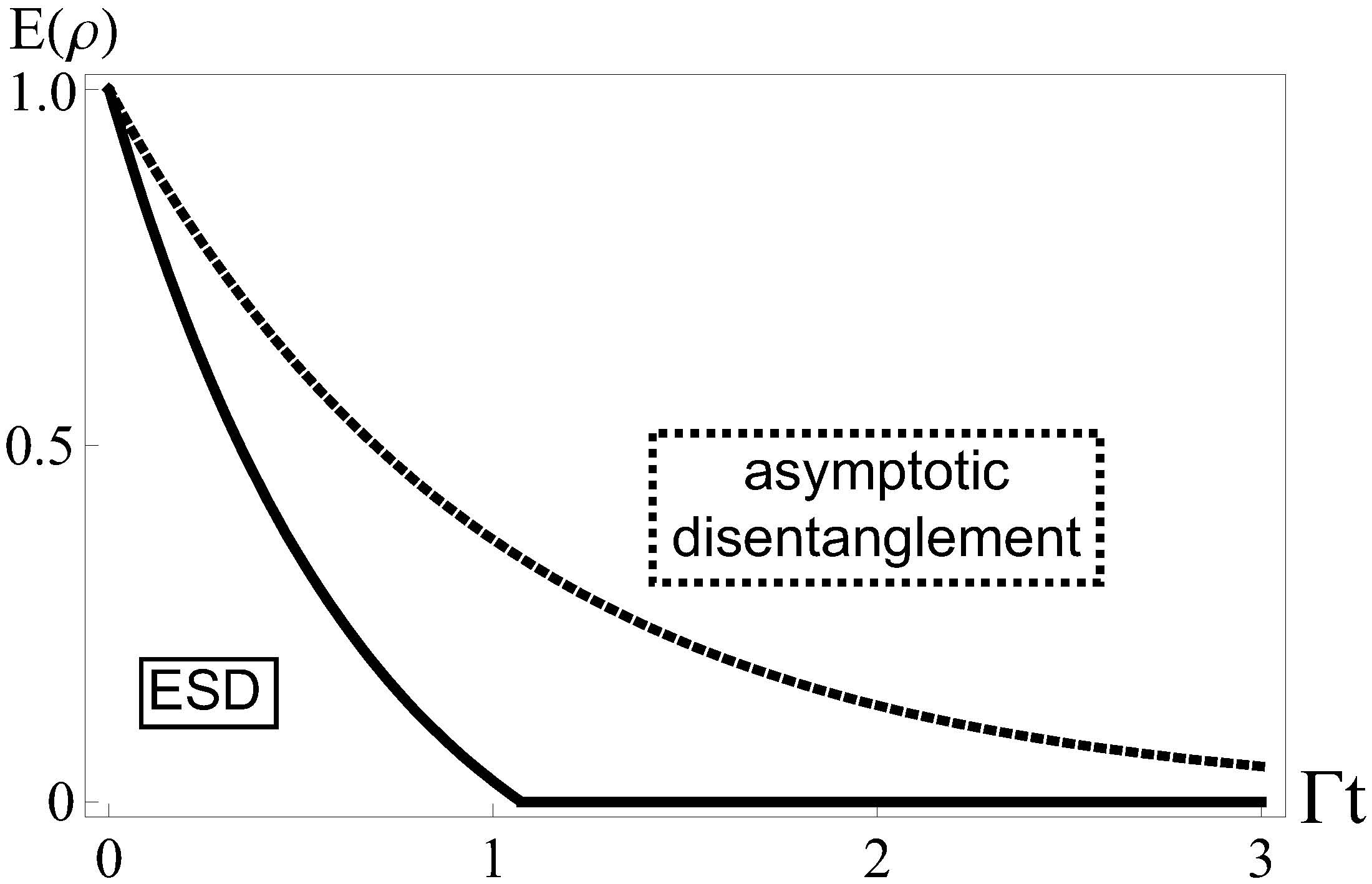}
	\caption{
	A generic plot of entanglement versus timescale, both in arbitrary units, showing the difference 
	between asympototic disentanglement (dashed line) and ESD (solid line), where the entanglement 
	goes abruptly to zero.  In both cases, quantum coherence decays exponentially.  This plot is 
	representative of the qualitatively different ways the entanglement of a system may evolve.
	}
	\label{fig02-asymptoticVsESD}
\end{figure}

Di$\acute{{\rm o}}$si's discovery of ESD was extended by Dodd and Halliwell \cite{DH04} and Dodd 
\cite{Dodd04} in a continuous variable system prepared in a Gaussian coherent Einstein-Podolsky-Rosen 
state in position and momentum, subjected to thermal environmental noise and negligible dissipation.  
The Wigner function $W(p_{1}, x_{1}, p_{2}, x_{2}) = W({\bf z}_{1}, {\bf z}_{2})$ 
evolving according to
\begin{equation}
\frac{\partial W}{\partial t} = 
- \frac{p_{1}}{m}\frac{\partial W}{\partial x_{1}}
- \frac{p_{2}}{m}\frac{\partial W}{\partial x_{2}}
+ D \frac{\partial^{2} W}{\partial p_{1}^{2}}
+ D \frac{\partial^{2} W}{\partial p_{2}^{2}} \ ,
\end{equation}
can be solved for the time-dependent solution
\begin{equation}
W_{t}({\bf z}_{1},{\bf z}_{2}) = \int d^{2}z_{1}'d^{2}z_{2}'
g({\bf z}_{1} - {\bf z}_{1}';\mathbb{A})g({\bf z}_{2} - {\bf z}_{2}';\mathbb{A})W_{0}'({\bf z}_{1}',{\bf z}_{2}') \ , 
\end{equation}
where the matrix
\begin{equation}
\mathbb{A} = D \ t
\left(
\begin{array}{cc}
 2 & t/m \\
 t/m & 2t^{2}/3m^{2}
\end{array}
\right)
\end{equation}
denotes the case of a dissipation-free free particle.  The Wigner function was shown to evolve 
into the separable form
\begin{equation}
W(p_{1},x_{1},p_{2},x_{2}) = \sum_{i} p_{i}W_{i}^{\rm A}(p_{1},x_{1})W_{i}^{\rm B}(p_{2},x_{2}) \ ,
\end{equation}
satisfying the continuous-variable Peres-Horode$\check{\rm c}$ki condition at a critical 
disentanglement time
\begin{equation}
t \geq 0.27 \left( \frac{\hbar m}{2 D}\right)^{1/2} \ ,
\end{equation}
where D is the dissipation constant and, therefore, exhibiting ESD.
Fig. \ref{fig02-asymptoticVsESD} shows the general relationship between asymptotic disentanglement
and Entanglement Sudden Death.

A discrete-variable model of spatially separated atoms in a cavity subjected to vacuum noise leading 
to spontaneous emission was next used in the search for additional examples of ESD \cite{YE04}.
Letting $H_{\rm at}$, $H_{\rm cav}$, and $H_{\rm int}$ represent the Hamiltonians for the atoms, 
cavities, and their interaction, respectively, the Hamiltonian for the total system is
$H_{\rm tot} = H_{\rm at} + H_{\rm cav} + H_{\rm int}$, with 
\begin{eqnarray}
H_{\rm at} &=& 
\frac{1}{2}\omega_{\rm A}\sigma_{\rm z}^{\rm A} + \frac{1}{2}\omega_{\rm B}\sigma_{\rm z}^{\rm B} \ , \\
H_{\rm cav} &=&
\sum_{\bf k}\omega_{\bf k}a_{\bf k}^{\dagger}a_{\bf k} + \sum_{\bf k}\nu_{\bf k}b_{\bf k}^{\dagger}b_{\bf k} \ , \\
H_{\rm int} = \sum_{\bf k}
\left( g_{\bf k}^{\ast}\sigma_{-}^{\rm A}a_{\bf k}^{\dagger} + g_{\bf k}\sigma_{+}^{\rm A}a_{\bf k} \right) &+& 
\sum_{\bf k}
\left( f_{\bf k}^{\ast}\sigma_{-}^{\rm B}b_{\bf k}^{\dagger} + f_{\bf k}\sigma_{+}^{\rm B}b_{\bf k} \right) \ ,
\end{eqnarray}
where $\left\{ g_{\bf k},f_{\bf k} \right\}$ are the system-environment coupling constants,  $\hat{a}(\hat{a}^{\dagger})$ and $\hat{b}(\hat{b}^{\dagger})$ the cavity field lower(raising) 
operators, $\sigma_{l}$ ($l= x, y, z$) 
are the Pauli operators, and $\sigma_{\pm}= \frac{1}{2}(\sigma_{x} \pm i\sigma_{y})$ 
are the atomic raising and lowering operators. 
In the two-qubit basis 
\begin{equation}
\ket{1}_{\rm AB} = \ket{++}_{\rm AB} \ , \ket{2}_{\rm AB} = \ket{+-}_{\rm AB} \ ,
\ket{3}_{\rm AB} = \ket{-+}_{\rm AB} \ , \ket{4}_{\rm AB} = \ket{--}_{\rm AB} \ 
\label{2qubitComputationalBasis}
\end{equation}
an important class of mixed states with $a \geq 0$, $d = 1 - a$, and $b = c = z = 1$,
is represented by
\begin{eqnarray}
\rho(t) = \frac{1}{3} \left(
\begin{array}{cccc}
 a(t) & 0        & 0    & 0   \\
 0    & b(t)     & z(t) & 0   \\
 0    & z(t)^{*} & c(t) & 0   \\
 0    & 0        & 0    & d(t)
\end{array}
\right) \ .
\end{eqnarray}
Its evolution given the operator-sum representation is
\begin{eqnarray}
\rho\left(t\right) = \sum_{\mu = 1}^{4}
K_{\mu}(t)\rho(0)K_{\mu}^{\dagger}(t) \ ,
\end{eqnarray}
where the operators representing amplitude damping noise, which satisfy the CPTP relations,
can be written
\begin{eqnarray}
K_{1} = 
\left(
\begin{array}{cc}
 \gamma_{\rm A} & 0\\
 0 & 1
\end{array}
\right) 
\otimes
\left(
\begin{array}{cc}
 \gamma_{\rm B} & 0\\
 0 & 1
\end{array}
\right) \ , \  
K_{2} = 
\left(
\begin{array}{cc}
 \gamma_{\rm A} & 0\\
 0 & 1
\end{array}
\right) 
\otimes
\left(
\begin{array}{cc}
 0 & 0\\
 \omega_{\rm B} & 0
\end{array}
\right) \ , \\
\nonumber \\
K_{3} = 
\left(
\begin{array}{cc}
 0 & 0\\
 \omega_{\rm A} & 0
\end{array}
\right) 
\otimes
\left(
\begin{array}{cc}
 \gamma_{\rm B} & 0\\
 0 & 1
\end{array}
\right) \ , \
K_{4} = 
\left(
\begin{array}{cc}
 0 & 0\\
 \omega_{\rm A} & 0
\end{array}
\right) 
\otimes
\left(
\begin{array}{cc}
 0 & 0\\
 \omega_{\rm B} & 1
\end{array}
\right) \ ,
\end{eqnarray}
where $\gamma_{\rm A(B)} = \gamma_{\rm A(B)}(t) = e^{-\Gamma_{\rm A(B)}t}$ 
characterizes the decay for subsystem A(B), described by the rate parameter 
$\Gamma_{\rm A(B)}$ and $\omega_{\rm A(B)} = \sqrt{1 - \gamma_{\rm A(B)}^{2}}$.  
In the Markov approximation, the two subsystems decohere at the same rate, so that 
$\Gamma_{\rm A} = \Gamma_{\rm B} = \Gamma$ with analogous relations,
$\gamma_{\rm A}(t) = \gamma_{\rm B}(t) = \gamma(t)$ and 
$\omega_{\rm A}(t) = \omega_{\rm B}(t) = \omega(t)$.
The concurrence is given by
\begin{equation}
C\big( \rho(t) \big) = \frac{2}{3}\max \left\{ 0, \gamma(t)^{2}f(t) \right\} \ ,
\end{equation}
with $f(t) = 1 - \sqrt{a(1 - a + 2\omega^{2} + \omega^{4}a)}$.  
The satisfaction of the inequality, $1 - a(1 - a + 2\omega^{2} + \omega^{4}a) \leq 0$, 
is a sufficient condition for concurrence to be zero.  Taking the case where $a = 1$, 
ESD occurs in the timescale
\begin{equation}
t_{\rm dis} = \frac{1}{\rm \Gamma}\ln\left[ \frac{2 + \sqrt{2}}{2}  \right] \ .
\end{equation}

Since the two-qubit system is important in many contexts, ESD was further developed in this 
simple, but interesting system.  The study of larger dimensional constituent subsystems of 
the bipartite state then followed, starting from the next discrete increments in the bipartite 
2$\times$3 and the 3$\times$3 cases, proceeding then to the arbitrary finite dimensional $d \times d$ case.
%before approaching the bipartite infinite-dimensional continuous-variable case described by 
%Wigner functions.  
As progress was made for increasing dimensions in the bipartite case, research 
also proceeded involving multipartite states of systems of an increasing numbers of qubits.

After the preliminary discovery of ESD using the simplifying assumptions of the Markovian 
approximation, further studies relaxed these conditions for a better undertstanding. 
Non-Markovian systems were considered that involve temperature effects, different sorts and 
combinations of noise, memory effects, as well as for different coupling strengths between 
various subsets of subsystems and environment.  Although ESD provides an intriguing avenue 
for mathematical research, for example, its geometrical interpretation, its ultimate importance 
lies with its physics.  To this end, its existence was proved in a wide variety of physical 
contexts such as in cavity QED, quantum optics, electrons on a solid state lattice, in 
superconducting systems like SQUIDs, and even in relativistic contexts.  After much theoretical 
development, experimental studies then confirmed ESD.  These further developments did not proceed 
in isolation; there is significant overlap among them since ESD is such a generic phenomenon.

%%%%%%%%%%%%%%%%%%%%%%%%%%%%%%%%%%%%%%%%%%%%%%%%%%%%%%%%%%%%%%%%%%%%%%%%%%%%%%%%%%%%%%%%%%%%%%%%%%%
\subsection{Theoretical developments}\label{Theoretical developments}

ESD is still a relatively recently discovered phenomenon, about which results have been 
theoretical ones.  These results are now surveyed in order of increasing
system complexity.

%%%%%%%%%%%%%%%%%%%%%%%%%%%%%%%%%%%%%%%%%%%%%%%%%%%%%%%%%%%%%%%%%%%%%%%%%%%%%%%%%%%%%%%%%%%%%%%%%%%
\subsubsection{Two qubits}\label{Two qubits}

The first paper showing ESD in a two-atom system under quantum vacuum noise leading to spontaneous 
emission \cite{YE04} of Yu and Eberly was followed by one examining the effects of ``classical'' 
noise \cite{YE06a}, that is, phase damping in a large class of two-qubit mixed states that often 
arise in physical contexts and includes the pure EPR-Bell states and the Werner mixed states.  
Two types of noise effects were considered, both global and multi-local, with the Hamiltonians 
\begin{eqnarray}
H_{\rm global} = - \frac{1}{2}\mu B(t)(\sigma_{z}^{\rm A} + \sigma_{z}^{\rm A}) \ , \\
H_{\rm multi-local} = - \frac{1}{2}\mu (b_{\rm A}\sigma_{z}^{\rm A} + b_{\rm B}\sigma_{z}^{\rm A}) \ , 
\end{eqnarray}
where $\mu$ is the gyromagnetic ratio and $B(t)$, $b_{\rm A}(t)$, and $b_{\rm B}(t)$ are the 
classical Markovian, white noise fields induced by an external magnetic field.  In the standard 
two-qubit computational of Eq. \ref{2qubitComputationalBasis}, the ``X-states'' are those that
can be written
\begin{eqnarray}
\rho_{\rm AB}(t) = 
\left(
\begin{array}{cccc}
 a &     0     & 0 & w \\
 0 &     b     & z & 0 \\
 0 &     z^{*} & c & 0 \\
 w^{*} & 0     & 0 & d
\end{array}
\right) \label{rhoX}\ ,
\end{eqnarray}
with the usual density matrix conditions such as Hermiticity, positive semi-definiteness, 
and normalization, $a + b + c + d = 1$.  When subject to such classical dephasing noise
describable in the operator sum representation similar to their earlier study \cite{YE04}, 
exponentially decaying off-diagonal terms appear for both the global and multi-local cases. 
Using concurrence as the entanglement measure, the critical time for complete disentanglement 
due to global dephasing noise was found to be
\begin{equation}
t_{\rm critical} = \frac{1}{2 \Gamma}\ln \frac{|w|}{\sqrt{bc}} \ ,
\end{equation}
where $\Gamma$ parameterizes the strength of dephasing and provided $b \neq 0$ and $c \neq 0$.  
For the multi-local dephasing case, disentanglement always occurs.  Thus, ESD may occur due 
to classical dephasing noise and so are not confined only to either quantum or amplitude 
damping noise.  The authors emphasized that this effect is independent of the specific 
entanglement measure considered. It is also independent of basis, because 
entanglement is so, once a decomposition of the system into subsystems has been chosen. 

An important later discovery concerned the non-additivity of weak noise influences in ESD.  Yu and Eberly 
considered the effects of weak dephasing noise and weak amplitude damping noise, each independently 
and when combined \cite{YE06b}, described by Hamiltonians and matrices similar to those 
used in their earlier studies of ESD \cite{YE04,YE06a}.  The concurrence was examined for 
the X-state density matrix as in Eq. \ref{rhoX}, with the following form, parameterized by the 
real quantity $\lambda$, chosen for simplicity and without loss of generality.
\begin{eqnarray}
\rho_{\lambda}^{\rm AB}(t) = \frac{1}{9}
\left(
\begin{array}{cccc}
 1 & 0       & 0       & 0 \\
 0 & 4       & \lambda & 0 \\
 0 & \lambda & 4       & 0 \\
 0 & 0       & 0       & 0
\end{array}
\right) \ .
\end{eqnarray}
At initial time, $C_{\lambda}(0) = 2\lambda/9$.  
For dephasing noise, 
$a(0) = a(t) = 1/9, d(0) = d(t) = 0, z(0) = \lambda /9 \rightarrow z(t) = \lambda / 9 \exp[-\Gamma_{2}t]$,
yielding a concurrence of
\begin{equation}
C_{\lambda}^{\rm ph.}(t) = (2\lambda/9)\exp[-\Gamma_{2}t] \ ,
\end{equation}
signifying asymptotic decay.  Under amplitude damping noise, one finds that
$z(0) = \lambda / 9 \rightarrow z(t) = \lambda / 9 \exp[-\Gamma_{1}t], a(0) = 1/9 \rightarrow a(t) = 1/9
\exp[-2\Gamma_{1}t], d(0) = 0   \rightarrow d(t) = 1/9 \omega_{1}^{4} + 8/9 \omega_{1}^{2}$, with 
$\omega_{1} = \sqrt{1 - \exp[- \Gamma_{1}t]}$ yielding a concurrence of
\begin{equation}
C_{\lambda}^{\rm amp.}(t) = 
\frac{2}{9}\left[ \lambda - \sqrt{\omega_{1}^{4} + 8\omega_{1}^{2}}\right]\exp[-\Gamma_{1}t] \ , 
\end{equation}
also signifying asymptotic decay.  In the above two expressions, $\Gamma_{1}$ and $\Gamma_{2}$ 
representing the strength of dephasing and amplitude damping, respectively.  When both dephasing 
and amplitude damping noises act simultaneously, their combined influence on the state was expected 
to be additive also.  However, the concurrence is given by
\begin{equation}
C_{\lambda}^{\rm ph. + amp.}(t) = 
2 \max\left\{ 0, \lambda e^{-\Gamma_{2}t} - \sqrt{\omega_{1}^{4} + 8\omega_{1}^{2}} \right\} \ 
\end{equation}
in that case, showing that the independent weak dephasing and amplitude damping noises are not 
additive, because the concurrence may go to zero in finite time and so that there is ESD.  
This study showed that the effects of noise on quantum systems, compared to noise acting on 
classical systems, are not just quantitatively different, but qualitatively so.  Yu and Eberly 
obtained further analytic results for these X-states in a follow-up analysis \cite{YE07c}, 
which also included state-equalizing noise and considered a subset of the X-states, the 
Werner states, given by
\begin{equation}
\rho_{\rm W} = 
\frac{1 - F}{3}\mathbb{I}_{\rm 4} + \frac{4F - 1}{3}\ket{\Psi^{-}}\bra{\Psi^{-}} \ ,
\end{equation}
with $1/4 \leq F \leq 1$, where $\ket{\Psi^{-}}$ is the Bell singlet state.  Among their results, 
it was shown that the Werner states subject to amplitude damping noise exhibited ESD only 
when equal to or less than a new critical fidelity $F_{\rm crit} \approx 0.714$, showing that 
the state was more robust against amplitude damping noise than to dephasing noise.  ESD was 
shown to exist for state-equalizing noise as well.  Another point here was that initial 
``local'' operations may act to protect from ESD, an important point related to decoherence-free 
subspaces.

The X-states were also considerd in an analysis by Ikram, Li, and Zubairy \cite{ILZ07}, 
who considered a two-qubit system interacting with a variety of dissipative noise 
environments (vacuum, thermal, and squeezed reservoirs) and showed that ESD occurs in 
all cases for squeezed reservoirs, as well as for thermal reservoirs with a nonzero number 
of photons.  However, for a vacuum reservoir where the mean thermal photon numbers are zero, 
$m = n = 0$, either ESD or asymptotic entanglement may occur as a function of initial 
conditions.  This particular study, similarly to many other studies of ESD, proceeded along 
the direction involving the consideration of different types of noise environments. However, 
some authors, have studied a bipartite state of two-level atoms 
subject to a controlled laser-induced Stark shift, for example, Abdel-Aty and Moya-Cessa \cite{AAMC07}.  Using the negativity
to measure entanglement, these authors showed that the Stark effect can produce either asymptotic 
entanglement or ESD.  As stated in Sec. \ref{Coherence}, alternative methods to study ESD 
may involve ``intrinsic'' decoherence, in contrast to the extrinsic decoherence from the 
open quantum systems viewpoint.

Further analysis of ESD in X-states was performed by Ban \cite{Ban08} from the perspective 
of the relationship of bipartite entanglement, as measured by the concurrence, and the 
phase correlation in bipartite X-states before and after the onset of ESD.  The phase 
correlation function of the X-state $\rho_{\rm AB}$ is given by
\begin{equation}
C(\theta_{\rm A}, \theta_{\rm B}) = 2 ( |w| + |z| ) \ ,
\end{equation}
%\begin{eqnarray}
%C(\theta_{\rm A}, \theta_{\rm B}) &=& 
%\left\langle  
%\left[ e^{-i(\hat{\phi} - \theta_{\rm A})} - \left\langle e^{-i(\hat{\phi} - \theta_{\rm A})} 
%\right\rangle \right]
%\otimes
%\left[e^{-i(\hat{\phi} - \theta_{\rm B})} - \left\langle e^{-i(\hat{\phi} - \theta_{\rm B})} 
%\right\rangle\right]
%\right\rangle \ ,
%\nonumber \\ &=& 2{\rm Re}\left[ x e^{-i(\theta_{\rm A} + \theta_{\rm B})}\right] + 
%2{\rm Re}\left[ y e^{-i(\theta_{\rm A} - \theta_{\rm B})}\right] \ ,
%\end{eqnarray}
%where $e^{-i(\hat{\phi} - \theta_{\rm i})}$ $(i = A, B)$ is the Pegg-Barnett phase operator 
%expressible in terms of the Pauli matrices as
%$e^{-i(\hat{\phi} - \theta_{\rm i})} = \sigma_{x}\cos(\theta_{i}) - \sigma_{y}\sin(\theta_{i})$.
where $|w|$ and $|z|$ are the off-diagonal elements of the X-state density matrix as in Eq. \ref{rhoX}.
Ban showed that before ESD occurs,
\begin{equation}
C\left(\rho_{\rm AB}(t < t_{\rm dis})\right) <  C\left(\theta_{\rm A}(t < t_{\rm dis}), \theta_{\rm B}(t < t_{\rm dis})\right).
\end{equation}
At the time of ESD,
\begin{equation}
C\left(\rho_{\rm AB}(t = t_{\rm dis})\right) =  C\left(\theta_{\rm A}(t = t_{\rm dis}), \theta_{\rm B}(t = t_{\rm dis})\right).
\end{equation}

%Since ESD may be caused by local noise, a natural question arises concerning the 
%relationship between local effects and state preservation against entanglement loss.  
%There exists a body of literature that examines this relationship focused on the search
%for decoherence free spaces (DFS), spaces of states that are unaffected by 
%environmental influences; the DFS states are very often entanglement states.  A 
%study of ESD modification by local operations therefore closely relates to DFSs, 
%already a large area of research of its own \cite{LW03}.  Let us briefly
%survey ESD protection in two-qubit systems to clarify this relationship before moving on
%to results regarding larger systems.

In other studies of the two-qubit systems, Jamr\'oz considered ESD in a two-qubit system 
due to spontaneous emission and showed that local unitary operations may indeed change the 
time for ESD onset from finite to infinite \cite{Jamroz06}.  Additionally, it was shown 
that for a large class of states including Werner states and pure states, a state may 
satisfy the two-qubit CHSH inequalities in finite time, even when ESD does not take place 
and the time for complete disentanglement is infinite.  Jordan, Shaji, and Sudarshan have 
discussed the effects of ``non-'' completely positive maps and their expression as local 
interactions that may increase or decrease the amount of entanglement present \cite{JSS07}.  
Furthermore, they discussed how noise influences that each separately disentangle 
asymptotically in time, when combined, may in finite time destroy all entanglement.  
Li, Chalaput, and Paraoanu consider a reservoir damping a two-qubit system with various 
sorts of noise such as: transverse, thermal squeezed, and longitudinal thermal \cite{LCP08}.  
The novelty of that study was that they were also subject to a continuous driving force that 
may act to decrease the time for ESD onset.  Thus, external local operations can both protect 
from ESD and may hasten it.  The development of ESD modification through local operations 
was studied by Rau, Ali, and Alber \cite{RAA08}, who showed that 
multi-local unitary operations may control the onset of entanglement sudden death by 
speeding it up or slowing it down to the point that it is never reached. Furthermore, they 
discovered that when such operations are performed on only one subsystem, it may change the 
rate of disentanglement, but cannot prevent it.  Dajka, Mierzejewski, and {\L}uczka considered
the control of ESD in ``non-'' Markovian systems \cite{DML08}.  Lastly, Maniscalco, Francica, 
Zaffino, Gullo, and Plastina even considered preserving entanglement and so preventing ESD 
through the quantum Zeno effect, which involves periodic strong projective measurements \cite{MFZ08}.

%%%%%%%%%%%%%%%%%%%%%%%%%%%%%%%%%%%%%%%%%%%%%%%%%%%%%%%%%%%%%%%%%%%%%%%%%%%%%%%%%%%%%%%%%%%%%%%%%%%
\subsubsection{Extensions: Higher dimensions or more subsystems}\label{Extensions}

The initial study of entanglement sudden death (ESD) in two-qubit systems was in the mean time quickly 
extended and complemented by research concerned with verifying and developing a better understanding 
of ESD in finite-dimensional systems by increasing the number of dimensions and subsystems, eventually 
approaching that of the continuous variables.  The existence of ESD in the next possible discrete-dimensional increment from the two-qubit system was first performed by Ann and Jaeger \cite{AJ08a} for a qubit-qutrit system using a methodology analogous to that of Yu and Eberly \cite{YE04} based on the the operator sum decomposition to describe multi-local dephasing noise operators 
acting on two-qubit systems and later for pairs of identical subsystems having any finite 
dimension. This line of research was carried out with an eye toward demonstrating
that effects that are pervasive in the universe can {\em in finite time} destroy entanglement in 
a broad range of systems first prepared in entangled states, not only those of the very simplest
case of pairs of two-level systems.

In these studies, dephasing noise was shown to induce a decay
in the off-diagonal elements of the density matrix of the system of interest in the form of an 
exponentially decay term and so decoherence only in the limit of infinite time 
while entirely destroying entanglement on a finite time scale. In this first example, 
the time-evolved qubit-qutrit density matrix was found to be
\begin{equation}
\hspace{8pt}\rho_{AB}(x,t)= \left(
\begin{array}{cccccc}
 \ \ \frac{1}{4} \ \ & \ \ 0 \ \ & \ \ 0 \ \ & \ \ 0 \ \ & \ \ 0 \ \ & \ \ x\gamma(t) \ \ \\
 0 & \frac{1}{8} & 0 & 0 & 0 & 0 \\
 0 & 0 & \frac{1}{8} & 0 & 0 & 0 \\
 0 & 0 & 0 & \frac{1}{8} & 0 & 0 \\
 0 & 0 & 0 & 0 & \frac{1}{8} & 0 \\
 x\gamma(t) & 0 & 0 & 0 & 0 & \frac{1}{4}
\end{array}
\label{ansatzWithZ} \right)\ ,
\end{equation}
where the diagonal elements $0 \leq x \leq \frac{1}{4}$ were chosen so that reduced 
subsystems exhibit no coherence and in order to satisfy Hermiticity, unitarity, and 
positive semi-definiteness, and $\gamma(t) = \exp [- \Gamma t]$ are the exponentially 
decaying decoherence factors with strength $\Gamma$.  The entanglement measured by the
negativity for this state was found to be 
\begin{equation}
N\left[ \rho_{\rm AB}(x, t) \right] = \max \left\{ 0, x\gamma(t) - \frac{1}{8} \right\} \ .
\end{equation}
Entanglement was here seen to be completely lost when $\gamma(t) = 1/8x$ at 
a time $t_{\rm dis} = 8x / \Gamma$, a finite time for the specified range of $x$ and for $\Gamma > 0$.
This results also showed that ESD is not restricted to identical subsystems. 

Considering the next higher dimensional increment, ESD in states of a qutrit pair 
was demonstrated by Che\c{c}ki\'nska and W\'odkiewicz in \cite{CW06,CW07} by 
examining the two-qutrit generalized Werner state
\begin{equation}
\rho_{\epsilon} = \frac{1 - \epsilon}{9}\mathbb{I}^{\rm A} \otimes \mathbb{I}^{\rm B} + 
\epsilon \ket{\Psi^{\rm AB}}\bra{\Psi^{\rm AB}} \ ,
\end{equation}
$0 \leq \epsilon \leq 1$ with entangled pure state and the pure entangled state 
\begin{equation}
\ket{\Psi} = \frac{1}{\sqrt{3}}\left( \ket{1^{\rm A}} \otimes \ket{1^{\rm B}} + 
\ket{2^{\rm A}} \otimes \ket{2^{\rm B}} + \ket{3^{\rm A}} \otimes \ket{3^{\rm B}}
\right) \ .
\end{equation}
In the two-qutrit basis 
$\left\{ \ket{00}, \ket{01}, \ket{02}, \ket{10}, \ket{11}, \ket{12}, \ket{20}, \ket{21}, \ket{22}\right\}$, 
the density matrix evolution is described by the operator-sum decomposition of the 
time-evolution operator with the following Kraus matrices,
\begin{eqnarray}
K_{0} = 
\left(
\begin{array}{ccc}
 1 & 0 & 0 \\
 0 & e^{-\frac{A_{1}t}{2}} & 0 \\
 0 & 0 & e^{-\frac{A_{2}t}{2}}
\end{array}
\right) , \  
K_{1} = 
\left(
\begin{array}{ccc}
 0 & \sqrt{1 - e^{-A_{1}t}} & 0 \\
 0 & 0 & 0 \\
 0 & 0 & 0
\end{array}
\right) , \ 
K_{2} = 
\left(
\begin{array}{ccc}
 0 & 0 & \sqrt{1 - e^{-A_{2}t}} \\
 0 & 0 & 0 \\
 0 & 0 & 0
\end{array}
\right) \ ,
\end{eqnarray}
where $A_{1}$ and $A_{2}$ are the Einstein coefficients representing the rate of amplitude damping 
and the Kraus matrices satisfy the CPTP relations.  The explicit time-dependent density matrix yields 
the explicit form of Caves and Milburn's separability condition \cite{CM00},
\begin{equation}
s(t) \equiv \frac{\epsilon}{8}\left( 2e^{\frac{1}{2}A_{1}t} + 2e^{-\frac{1}{2}A_{2}t} + 
2e^{-\frac{1}{2}(A_{1}+A_{2})} + e^{-A_{1}t} + e^{-A_{2}t} \right) \leq \frac{1}{4} \ ,
\end{equation}
and so
\begin{eqnarray}
s(t = 0) = \epsilon \leq \frac{1}{4} \ ,
\end{eqnarray}
indicating separability; for the maximum value $\epsilon = 1$, separability was seen to arise in finite time.

The demonstration of ESD in a bipartite system of a pair of identical subsystems of arbitrarily large finite dimensions was then taken by Ann and Jaeger \cite{AJ07b}, by considering the case for which a mixed state entanglement 
measure exists for arbitrary $d > 2$, namely, the $d \times d$ isotropic states.  The isotropic 
states are those that are invariant under $U \otimes U^{*}$ transformations and are of the form
\begin{equation}
\rho_{\rm iso}(d) = \left( \ \frac{1 - F}{d^{2} - 1} \right) I_{d^{2}} + \left(\ \frac{F d^{2} - 1}{d^{2} - 1} \right) P(\ket{\Psi(d)}) \ ,
\end{equation}
defined in Sec. \ref{Entanglement measures}.  This state is separable when 
$F\left(\rho_{\rm iso}(d),P(\ket{\Psi})\right)\leq F_{\rm critical}(d)\equiv d^{-1}$, 
according to Terhal and Vollbrecht's expression for entanglement of formation discussed 
in Sec. \ref{Entanglement measures}.  Specifically, the isotropic states must satisfy 
``both'' of the following conditions,
\begin{eqnarray} 
{\rm (i)}  & &  F\left(\rho_{\rm iso}(d),P(\ket{\Psi(t = 0)})\right)         > F_{\rm critical}(d) \ , \ \ \ \ {\rm and} \\
{\rm (ii)} & &  F\left(\rho_{\rm iso}(d),P(\ket{\Psi(t < \infty)})\right) \leq F_{\rm critical}(d) 
\end{eqnarray}
for some finite time $t$, for there to be ESD.  These conditions were both shown to be satisfied 
for an isotropic state subject to depolarizing noise; an initially entangled state becomes 
separable in finite time despite the persistence of state coherence.  This demonstrated 
ESD in $d \times d$ isotropic states for arbitrary finite dimensions $d > 2$.  A later study 
also examined entanglement evolution of isotropic states \cite{TDB08}, but by making
use of the G-concurrence \cite{Gour05}, which involves the multiplication of ``all'' of a state's 
Schmidt coefficients.  Therefore, if a single Schmidt coefficient vanishes, the G-concurrence 
vanishes for a $d$-level system.  That study specifically showed that a $d \times d$ isotropic state 
subject to depolarizing noise, parameterized by a strength $\Gamma$, undergoes ESD at the $k^{th}$ 
level ($2 \leq k \leq d$) in a time $t_{k} = \ln[(d^{2} - 1)/(dk - d - 1)]/(2d\Gamma)$.

As the dimensions of the bipartite case are indefinitely incremented, one approaches the case of 
infinite discrete dimensions and the continuous variables case.  The work of Di$\acute{{\rm o}}$si 
\cite{Diosi03}, Dodd and Halliwell \cite{DH04}, and Dodd \cite{Dodd04} discussed in Sec. \ref{First discoveries} 
on continuous-variable states was among the first to demonstrate ESD under local noise. Further 
work is still required to understand better and clarify the nuances of the relationship between the 
finite and infinite dimensional cases.

Investigations of ESD in systems of larger numbers of constituents has also very recently been pursued,
for example, in a study involving ESD in a three-qubit system composed of three atoms A, B, 
and C each interacting only with their own isolated cavities a, b, and c was recently 
performed \cite{MFL08}.  The total Hamiltonian for the system is composed of three Jaynes-Cummings 
Hamiltonians,
\begin{eqnarray}
H = \sum_{i=A,B,C} \frac{1}{2} \sigma_{z}^{i} + 
\omega a_{i}^{\dagger}a_{i} + g(a^{\dagger}\sigma_{-}^{i} + a\sigma_{+}^{i}) \ ,
\end{eqnarray}
where $\omega_{0}$ and $\omega$ are the atomic and cavity frequencies, respectively.
For initial tripartite GHZ and W states, the authors were forced, in fact, only to study 
negativity for bipartitions of the system: A-BC, a-bc, A-bc, Ab-c, aB-C, and a-BC, because 
negativity is not appropriate for mixed tripartite states.  At the bipartite level, ESD was 
shown to exist for this set of bipartitions for the tripartite GHZ state, but not for the W 
state.  Since these states exhibit a high degree of symmetry, this study strongly suggests, 
but does not definitively confirm genuine tripartite ESD, because a mixed-state entanglement 
measure does not exist, despite the authors' claims to have done so.

The entanglement properties of multipartite GHZ-states and W-states were earlier studied 
using the specific form of the multipartite concurrence in Eq. \ref{nConcurrence} \cite{CMB04}.
The time evolution was studied using the master equation
\begin{eqnarray}
\frac{d\rho_{k}}{dt} = \sum_{k = 1}^{N}
\underbrace{(\mathbb{I} \otimes \cdots \otimes \mathbb{I})}_{k-1} \otimes L_{k} \otimes
\underbrace{(\mathbb{I} \otimes \cdots \otimes \mathbb{I})}_{N - k}\rho \ ,
\end{eqnarray}
where $\rho$ and $\rho_{k}$ are the full system and reduced density matrices, respectively, 
and $L_{k}$ are the Lindblad operators representing multi-local weak Markovian noise,
\begin{equation}
L_{k}\rho = \sum_{i} = \frac{\Gamma_{i}}{2}
(2a_{i}\rho a_{i}^{\dagger} - a_{i}^{\dagger}a_{i}\rho - \rho a_{i}^{\dagger}a_{i}) \ ,
\end{equation}
where $\Gamma_{i}$ represents the strength of the noise and the creation(annihilation)
operators as defined in Sec. \ref{First discoveries}.  The concurrence was computed 
numerically for initial three-qubit GHZ-states and W-states; it was found that despite 
coherence only asymptotically decaying for the zero-temperature case, entanglement 
vanishes in finite time for non-zero temperatures.  

In general, studies suggest that entanglement under environmental influences is more robust 
for multipartite W-states than for the GHZ-states for a wide variety of situations such as 
scaling in the number of subsystems, temperature, and time.  Multipartite ESD due to multi-local 
noise was also addressed by Aolita, Chaves, Cavalcanti, Ac\'\i n, and Davidovich 
\cite{ACC08}.  These investigators considered a class of multipartite states in which 
the entanglement calculation was considered in the most balanced partition, and therefore
effectively reduced to a four-dimensional problem and quantifiable by the negativity.  They 
found that even though the ESD time is an increasing function with the number of qubits, 
the more important timescale for entanglement to be arbitrarily small is instead a 
monotonically decreasing function with the number of qubits.

The existence of ESD for ranges of initial states in such a wide variety of contexts 
strongly suggest that ESD is a generic phenomenon in all quantum systems in specific 
classes of states.  Given that, in general, systems may enter any of a broad range of
states, it appears unlikely that entanglement will be found to persist in nature, as
intuitively foreseen by Schr$\ddot{\rm o}$dinger in his cat paper \cite{Schrodinger35}.

%%%%%%%%%%%%%%%%%%%%%%%%%%%%%%%%%%%%%%%%%%%%%%%%%%%%%%%%%%%%%%%%%%%%%%%%%%%%%%%%%%%%%%%%%%%%%%%%%%%
\subsubsection{Geometry}\label{Geometry}

In addition to its extension to a greater number of dimensions, the study of ESD has also been 
developed from a geometric perspective.  Entanglement evolution and ESD in a two-qubit system 
are examined by Cunha \cite{Cunha07} and Drumond and Cunha \cite{DC08}.  State evolutions were 
shown to have different asymptotic states: (i) when the asymptotic state is separable, ESD may 
exist as a feature of the time evolution; (ii) when the asymptotic state exists on the border 
of the entangled and separable states, both sudden death and asymptotic disentanglement may exist;
(iii) when the asymptotic state is entangled, in which case entanglement sudden birth may be a 
feature of evolution; and (iv) there may exist a multiple asympototic states, in which case some 
combination of all three of the previous may exist. 

Paz and Roncaglia considered this geometric interpretation of asymptotic state evolution in an 
explicit physical model by examining two resonant quantum harmonic oscillators coupled to a 
mutual environment using the quantum Brownian motion master equation \cite{PR08}. Taking into 
account model parameters including temperature, ohmicity, and squeezing factors, they showed
the existence of and mapped out the boundaries of the three phases in this system: sudden death, 
sudden death and sudden birth cycles, and no sudden death.

\begin{figure}[htbp]
	\centering
	\includegraphics[width=0.50\textwidth]{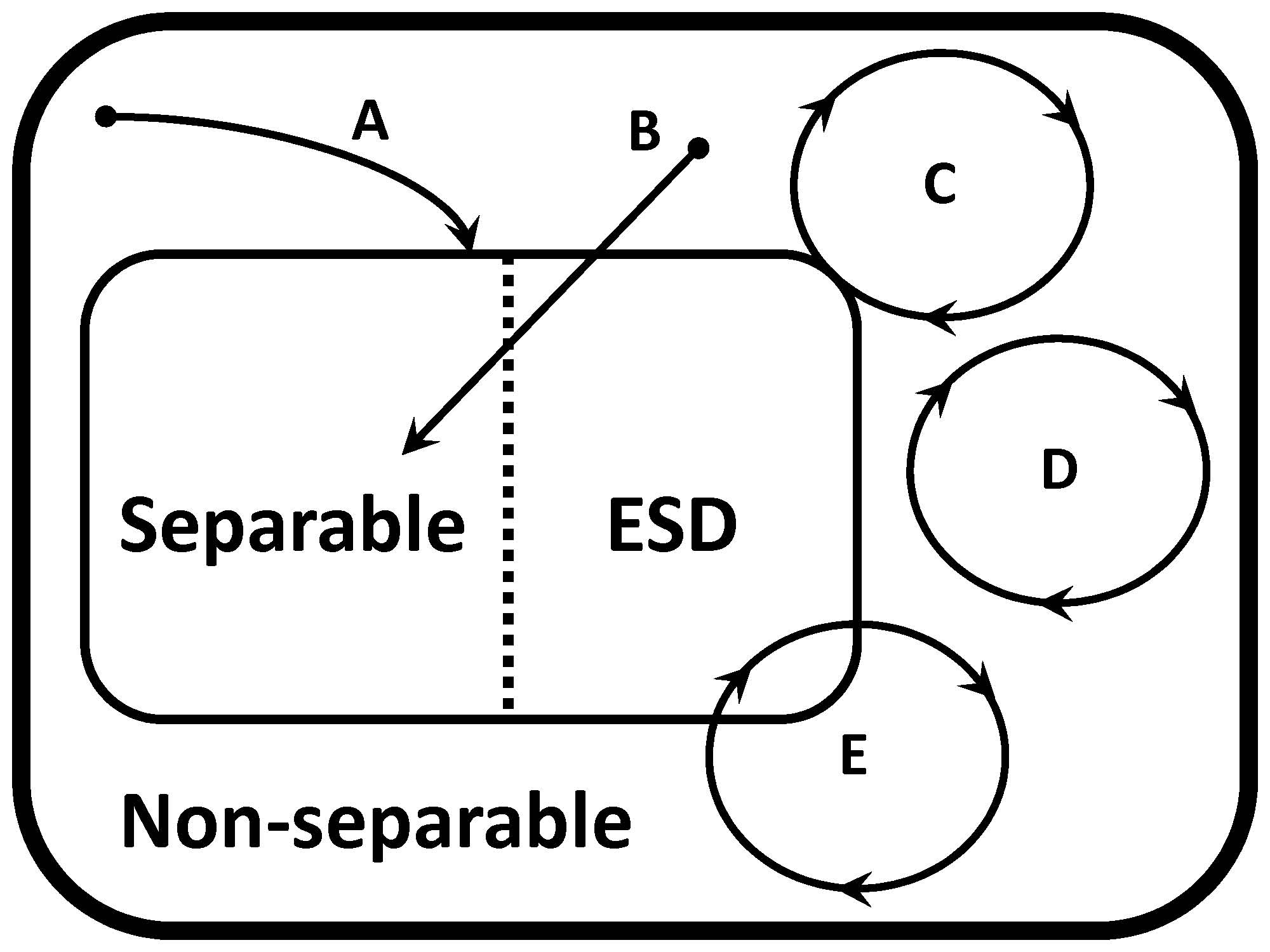}
	\caption{A geometric picture of how initially entangled, non-separable states may evolve in time
	to final asymptotic states.
	(A) Asymptotic disentanglement: a separable state is approached asymptotically in time.
	(B) Entanglement sudden death (ESD): entanglement is completely and abruptly lost in finite time.
	(C) Coexistence of multiple asymptotic states, entangled and/or separable.
	(D) Entanglement is maintained for the entire state evolution.
	(E) Entanglement sudden deaths and births: entanglement may abruptly reappear after ESD, furthermore, 
	this may occur in cycles.
}
	\label{fig03-ESDregions}
\end{figure}

The degree of entanglement was considered to be the distance between a quantum state and the 
nearest separable state was considered by Shimony \cite{Shimony95}, who defined the distance as
\begin{equation}
E_{\rm S}(\ket{\Psi}) = \frac{1}{2}\min \left\| \ket{\Psi} - \ket{\Xi} \right\| \ ,
\end{equation}
where $\ket{\Xi}$ is a normalized separable state and the minimization is taken over all such states.
%using a natural distance function between two states $\ket{\psi_{1}}$ and $\ket{\psi_{2}}$ 
%as the Hilbert space angle defined in terms of the overlap between the two states,
%$D(\ket{\psi_{1}},\ket{\psi_{2}}) = \phi = \cos^{-1}\left( |\bra{\psi_{1}}\ket{\psi_{2}}|\right)$.
The notion of entanglement distance was extended by Yu and Eberly \cite{YE07b}, who consider 
the notion of an entanglement distance measuring the distance between a given state and the 
boundary of separable states with entangled states and defined by the argument of the 
concurrence function $\Lambda$.  $\Lambda > 0$, $\Lambda = 0$, and $\Lambda < 0$ represent, respectively, 
the cases where the state is entangled, purely separable, and ``both'' mixed and separable.  
For example, for states that stay only within a separable or an entangled region, and do not 
become entangled or lose entanglement, respectively, $\Lambda(t)$ does not change sign.  
In contrast, for those states exhibiting entanglement sudden death and birth, $\Lambda(t)$ 
may change sign in time.  By using this measure to examine the entanglement time evolution, 
a rich understanding from a different perspective is obtained.
Fig. \ref{fig03-ESDregions} provides a visual representation of asympototic state evolution.

Although such a geometric perspective for two-qubit states is illuminating, the results of this 
analysis pose further questions as to the proper geometric description of other entanglement 
related concepts such as bound entanglement, higher-dimensional entanglement, multipartite 
entanglement, as well as ESD for nonlinear systems.  Furthermore, these geometric ideas ought
to be describable in terms of previously developed geometric characterizations of entanglement, 
such as that of the Minkowskian length associated with the Stokes tensor \cite{JSST03}.

%%%%%%%%%%%%%%%%%%%%%%%%%%%%%%%%%%%%%%%%%%%%%%%%%%%%%%%%%%%%%%%%%%%%%%%%%%%%%%%%%%%%%%%%%%%%%%%%%%%
\subsubsection{Non-Markovian dynamics}\label{Non-Markovian dynamics}

Many analyses of Entanglement Sudden Death were done assuming the Markovian approximation, where 
calculations are relatively clean and simple analytical results may be obtained, leaving
a wide range of less pervasive phenomena to be discovered.  In particular, the case of non-Markovian noise 
was rarely considered until recently.  When the simplifying assumptions of the Markovian approximation are 
relaxed, many interesting phenomena arise that aren't possible under that approximation arise, 
such as backreactive effects, memory effects, and entanglement creation.

The studies so far performed all proceeded in the same following way.  First, the Hamiltonians 
are specified for the system $H_{\rm sys}$ and environment $H_{\rm env}$, along with the 
interaction Hamiltonian $H_{\rm int}$ that couples them to each other.  Second, an initial class 
of states are considered, often with other important initial conditions specified.  Third, the 
solution to the time-evolved state is found.  Lastly, entanglement dynamics are analyzed according 
to the measures introduced in Sec. \ref{Entanglement measures}.  The reader is referred to the 
appropriate cited publication for specific details.  When needed, the total Hamiltonian describing 
the system-environment dynamics is given using the relations defined at the beginning of Sec. 
\ref{First discoveries} along with the initial conditions. 

Anastopoulos, Shresta, and Hu considered a two-qubit system consisting of atoms coupled to a common 
electromagnetic field and derived the non-Markovian master equation for each of the atoms \cite{ASH07}.  
Their approach is analogous to the study of \cite{YE04}, but in the non-Markovian regime.  The 
total Hamiltonian of the atom--field system is given by $H_{\rm tot} = H_{\rm sys} + H_{\rm int}$, with
\begin{eqnarray}
H_{\rm sys} = \hbar \sum_{k}\omega_{k}\hat{a}_{k}^{\dagger}\hat{a}_{k} + \hbar \omega_{0}
(\sigma_{+}^{\rm A}\sigma_{-}^{\rm A} + \sigma_{+}^{\rm B}\sigma_{-}^{\rm B}) \ \ \ \ \ \ \  \\
H_{\rm int} = \hbar \sum_{k} g_{k}
\left( 
\hat{a}_{k}^{\dagger}(e^{-i{\bf k}\cdot{\bf r}}/2\sigma_{-}^{\rm A} +  e^{+i{\bf k}\cdot{\bf r}}/2\sigma_{-}^{\rm B})
+         \hat{a}_{k}(e^{+i{\bf k}\cdot{\bf r}}/2\sigma_{+}^{\rm A} +  e^{-i{\bf k}\cdot{\bf r}}/2\sigma_{+}^{\rm B})
\right) \ \ \ \ \ \ \ \ 
\end{eqnarray}
where $g_{k} = \lambda / \sqrt{\omega_{k}}$ and $\lambda$ is the system-environment coupling strength.  
The authors considered two classes of states, 
$\ket{A} = \sqrt{p}\ket{11} + \sqrt{1-p}\ket{00}$ 
with $0 \leq p \leq 1$, 
and $\ket{B} = x \ket{+} + (1-x) \ket{-}$, 
with $\ket{+} = (\ket{01} + \ket{10})/\sqrt{2}$ 
and  $\ket{-} = (\ket{01} - \ket{10})/\sqrt{2}$.
It was shown that, for class B states, the results are identical to those for the Markovian 
case.  However, for class A states, entanglement behavior is dependent on the distance between 
the two atoms.  Furthermore, the entanglement exhibited sudden birth and death cycles.
Therefore, we see that the entanglement evolution is dependent on initial conditions.  

Bellomo, Franco, and Compagno considered a method to characterize dynamics in the case of $N$ 
subsystems, each influenced different sorts of local noise \cite{BFC07}.  This general model 
was also applied specifically to a non-Markovian two-qubit model in a zero-temperature 
environment, with the Hamiltonian  
\begin{equation}
H_{\rm tot} = 
\omega_{0}\sigma_{+}\sigma_{-} + \sum_{k}\omega_{k}b_{k}^{\dagger}b_{k} + (\sigma_{+}B + \sigma_{-}B^{\dagger}) \ ,
\end{equation}
where $B = \Sigma_{k}g_{k}b_{k}$ with $g_{k}$ being the coupling strength.  The remaining  
studies discussed in this section used a similar model.  Despite the lack of direct 
interaction between the two qubits, it is shown that their entanglement may exhibit cycles 
of death and birth.  

Dajka, Mierzejewski, and {\L}uczka \cite{DML07} considered the exact non-Markovian dynamics 
of a separated two-qubit system, with just one of them subject to local dephasing noise 
described by the Hamiltonian is given by
\begin{equation}
H_{\rm tot} = \sigma_{z}^{\rm A} + \sigma_{z}^{\rm B} + 
\sum_{k=1}^{\infty}g_{k}(\hat{a}_{k}^{\dagger} + \hat{a})\sigma_{z}^{\rm B} +
\sum_{k=1}^{\infty}\omega_{k}\hat{a}_{k}^{\dagger}\hat{a}_{k} \ .
\end{equation}
The reduced subsystems were analyzed for arbitrary subsystem-environment coupling strengths 
and environmental fluctuation frequencies.  The following different entanglement behaviors
were formal: ESD exists for the subohmic and ohmic environments, but there is only asymptotic 
disentanglement for a zero-temperature superohmic environment.  In a separate study of a 
non-Markovian two-qubit system, Cao and Zhen studied ESD at zero temperature as a function 
of coupling strength and they found that if the coupling is sufficiently strong, ESD occurs, 
even at zero temperature \cite{CZ08}.

Liu and Goan recently studied a continuous variable system in a thermal environment subject 
to non-Markovian dynamics \cite{LG07}.  Two types of joint system-environment coupling models 
were considered, the first one having each oscillator coupled to its own environment and the 
second both oscillator coupled to a common environment.  This model is analogous to \cite{YE03} 
for multi-local noise and collective noise, but using oscillators instead of qubits.  Entanglement 
was quantified by the logarithmic of the negativity.  Unsurprisingly, behavior was found to be 
dependent on many factors including the initial state, inter-subsystem coupling, subsystem-environment 
coupling, and the influence of multi-local or global noise.  ESD was demonstrated to exist for 
some of these situations and not for others.  Another analysis of a similar comprehensive 
nature as this was undertaken by Bellomo, Franco, and Compagno, in which they studied the 
entanglement dynamics of a two-qubit Werner state as a function of purity, initial state 
entanglement, and the extent of the non-Markovian noise character with similar results \cite{BFC08}.

 Many models exist with even more complex entanglement 
evolution characteristics, for example, complete loss of entanglement near quantum phase 
transitions \cite{Hamdouni08}.  Although this case would not strictly fit the classification 
of ESD, which depends on time, it may conform to the general idea of ESD if the quantum phase
transition itself could be parameterized as a function of time.  

%%%%%%%%%%%%%%%%%%%%%%%%%%%%%%%%%%%%%%%%%%%%%%%%%%%%%%%%%%%%%%%%%%%%%%%%%%%%%%%%%%%%%%%%%%%%%%%%%%%
\subsubsection{Energy and temperature}\label{Energy and temperature}

Several studies of ESD have also been carried out involving its relationship to energy and temperature,
providing another perspective on the phenomenon.

Cui, Li, and Yi considered a two-qubit system coupled to an electromagnetic field and 
analyzed the relationship between entanglement evolution and energy transfer when subject to 
both open systems and closed systems dynamics \cite{CLY07}. Considering the (open systems) dynamics 
involving amplitude damping and dephasing, they obtained results confirming previous 
studies on ESD \cite{YE04}.  The Hamiltonian
\begin{equation}
H = \frac{\omega}{2}(\sigma_{\rm A}^{z} + \sigma_{\rm B}^{z}) + \frac{g}{2}\sigma_{\rm A}^{x}\sigma_{\rm B}^{x} \ ,
\end{equation}
was used to model closed systems dynamics.  They found close relationships between 
entanglement and energy transfer, even suggesting that entanglement, expressed by the concurrence, 
could be a function of the energy transfer, $C[\rho(t)] = \max\left\{ 0, f(E) \right\}$, and that 
entanglement exists above a critical energy $E_{\rm crit}$ but not below it.  Furthermore, energy 
and system evolution may be different after the onset of ESD.  A similar study was performed by 
Li, Fu, and Liang \cite{LFL07} and showed that an EPR state of two-atoms in a cavity disentangles 
faster for a larger initial mean photon number $\bar{n}$ and that the maximum concurrence corresponds 
to minimum energy of the system.  However, entanglement and energy are neither mapped in a one-to-one 
fashion nor evolve at the same rate.

The quantitative relationship between disentanglement and system-environment energy transfer 
has been developed in more detail by Yu \cite{Yu07}.  A two-qubit system in the form of the 
X-state in Eq. \ref{rhoX} with $w = 0$ evolves according to the
Hamiltonian $H_{\rm tot} = H_{\rm sys} + H_{\rm int} + H_{\rm env}$, with 
\begin{eqnarray}
H_{\rm sys} = \frac{1}{2}E_{\rm A}\sigma_{z}^{\rm A} + \frac{1}{2}E_{\rm B}\sigma_{z}^{\rm A} \ , \\
H_{\rm env} = \sum_{\lambda} \omega_{\lambda}a_{\lambda}^{\dagger}a_{\lambda} + 
\sum_{\lambda} \nu_{\lambda}b_{\lambda}^{\dagger}b_{\lambda} \ , {\rm and} \ \ \\ 
H_{\rm int} = \sum_{\lambda}(f_{\lambda}^{\ast}\sigma_{\rm A}a_{\lambda}^{\dagger} +  
f_{\lambda}\sigma_{\rm A}^{\dagger}a_{\lambda}) + 
\sum_{\lambda}(g_{\lambda}^{\ast}\sigma_{\rm A}b_{\lambda}^{\dagger} +  
g_{\lambda}\sigma_{\rm B}^{\dagger}b_{\lambda}) \ .
\end{eqnarray}
An explicit upper bound for the amount of energy transfer required for complete 
disentanglement of the initial state was found: 
\begin{equation}
|\Delta {\rm E}| \leq \frac{E_{\rm A}(2\bar{n} + 1)}{2[(2\bar{n} + 1)^{2} + 2\bar{n}(\bar{n} + 1)]} \ ,
\end{equation} 
where $\bar{n}$ is the mean number of thermal quanta in the reservoir.

An avenue of research differing from all other open quantum systems studies involving 
extrinsic decoherence was undertaken by Silman, Machnes, Shnider, Horwitz, and Belenkiy, 
where ``intrinsic'' decoherence is considered by the addition of a stochastic term to the 
Schr$\ddot{{\rm o}}$dinger equation \cite{SMS08}.  The evolution of the initial singlet 
state $\ket{\Psi} = (\ket{01} - \ket{10})/\sqrt{2}$ was modeled according to the 
Schr$\ddot{{\rm o}}$dinger equation in the stochastic reduction framework in the energy basis,  
\begin{equation}
{\rm d}\ket{\psi(t)} = -i\hat{H}\ket{\psi(t)}{\rm d}t - 
\frac{1}{8}\sum_{i=A,B}\zeta_{i}^{2}(\hat{H}_{i} - H(t))^{2}\ket{\psi(t)}{\rm d}t +
\frac{1}{2}\sum_{i=A,B}\zeta_{i}    (\hat{H}_{i} - H(t))    \ket{\psi(t)}{\rm d}W_{i}(t) \ ,
\end{equation}
where $\ket{\Psi(t)}$ is the normalized state vector, $W(t)$ is a Wiener process, $\zeta$ is 
a parameter specifying the reduction timescale, and ${\rm d}W_{i}(t){\rm d}W_{j}(t) = \delta_{ij}{\rm d}t$.
After averaging over the noise fields, the state density matrix was obtained by the Lindblad equation, 
\begin{equation}
\frac{{\rm d}}{{\rm d}t}E[\rho(t)] = -i[\hat{H}, E[\rho(t)]] - \sum_{i=A,B}
\frac{\zeta_{i}^{2}}{8}[\hat{H}_{i},[\hat{H}_{i},E(\rho(t))]] \ .
\end{equation}
Using the negativity to quantify entanglement, they determined the conditions for either 
ESD or asymptotic disentanglement to take place as a function of such model parameters as the 
stochastic reduction rate and the direction of the measurement device relative to the motional 
direction of the particles of the singlet.

Many studies of the sudden death of global state properties not only assumed the Markovian 
approximation, but also zero temperature.  A natural question is how to extend the results
of these studies to the case of finite temperature $T > 0$.  
In order to address this quesetion, Jak\'obczyk and 
Jamr\'oz considered a two-qubit system coupled to a finite-temperature bath evolving according 
to the following master equation, 
\begin{equation}
\frac{d\rho}{dt} = 
\frac{1}{2}\Gamma_{\uparrow} \left\{[\sigma_{+},\rho \sigma_{-}]+[\sigma_{+}\rho,\sigma_{-}] \right\} + 
\frac{1}{2}\Gamma_{\downarrow} \left\{[\sigma_{-},\rho \sigma_{+}]+[\sigma_{-}\rho,\sigma_{+}] \right\} \ ,
\end{equation}
where $\Gamma_{\uparrow}   = \gamma_{0}n(\omega_{0})$,
$\Gamma_{\downarrow} = \gamma_{0}\left( 1 + n(\omega_{0})\right)$, 
$n(\omega_{0}) = \frac{1}{e^{\beta \omega_{0}} - 1}$, and 
$\beta = \frac{1}{T}$ \cite{JJ04}.
All initially entangled states in a noisy cavity were found to disentangle on a finite time scale
that is an increasing function of the initial state entanglement.  Furthermore, they showed 
the onset of loss of standard Bell inequality violation occurs on a shorter timescale.
Al-Qasimi and James in a similar analysis considered a two-qubit system initially prepared 
in the X-states and coupled to some finite temperature exhibit ESD \cite{AQJ08}.  Notably, 
these classes exhibiting ESD contain states that were robust against ESD in the zero 
temperature regime.

These studies suggest that ESD is not only a novel and counterintuitive result, but it is more 
likely than long-lived entanglement in typical contexts, which involve energy transfer 
and temperature effects.  Practical tasks such as quantum error correction protocols necessarily 
need also to account for energy transfer and temperature-induced effects that degrade entanglement 
and coherence themselves in addition to the other sorts of phase and amplitude damping errors. 

%%%%%%%%%%%%%%%%%%%%%%%%%%%%%%%%%%%%%%%%%%%%%%%%%%%%%%%%%%%%%%%%%%%%%%%%%%%%%%%%%%%%%%%%%%%%%%%%%%%
\subsubsection{Baths: spins and oscillators}\label{Baths: spins and oscillators}

Many studies of ESD consider a central system of interest composed of spins and/or oscillators 
surrounded by and interacting with a bath of spins and/or oscillators. In this section, the case of 
spins is mainly considered.  The dynamics for 
the central system are then solved for, either before or after tracing over the environment
variables of the state, at which point the coherence and entanglement dynamics are examined.  
They also usually involve 
changing subsystem-environment coupling strengths or distance scales and other aspects of 
the model environment.  There is often reference to different environments according to their 
spectral properties.  It is therefore useful to recall that the environmental spectral function
in the thermodynamic limit is 
$J( \omega) = \lambda \omega^{1 + \mu}\exp [-\omega/\omega_{\rm crit}]$, where $\omega_{\rm crit}$ 
is the environmental cutoff frequency and $\lambda$ is the qubit-environment coupling strength; 
the sub-ohmic, ohmic, and super-ohmic case corresponds to $\mu \in (-1,0)$, $\mu = 0$, and 
$\mu \in (0, \infty)$, respectively ({\it cf.} \cite{LCD87}).  

Roszak and Machnikowski study ESD in a two-qubit system subject to pure dephasing influences
from a super-ohmic environment and the effect of inter-qubit distance \cite{RM06}.  The two-qubits and
environment represent excitonic qubits and phonons, respectively.  For the initial 
states $\ket{\psi_{0}^{(1)}} = \frac{\ket{00} + \ket{01} + \ket{10} - \ket{11}}{2}$ and
$\ket{\psi_{0}^{(2)}} = \frac{\ket{01} - \ket{10}}{\sqrt{2}}$, the entanglement dynamics 
were analyzed using concurrence.  ESD was shown to exist as a function of different 
initial states and temperatures.  Additionally, inter-qubit distance was shown to be 
important since ESD only occurs for spatially separated states.  

The importance of distance, mentioned by Schr$\ddot{\rm o}$dinger \cite{Schrodinger35}, 
between subsystems in relation to entanglement was carefully considered by Cormick and Paz, 
who studied a two-qubit system locally coupled to an environmental XY-spin chain \cite{CP08}.  
After obtaining an exact solution of the system of interest, they analyze how decoherence 
and disentanglement behave as a function of distance and coupling strength.  
For strong system-environment couplings, quasiperiodic entanglement sudden deaths and 
sudden births as a function of the inter-qubit distance were found.
%%%%%
Jing, L$\ddot{\rm u}$, and Yang \cite{JLY07} consider a two-qubit Bell state where one qubit 
is not coupled to the environment and the other interacts with a Heisenberg XY spin bath in 
thermal equilibrium.  In one case, the bath contains a finite number of spins ($N=40$) and 
in another case an infinite number of spins (i.e. $N \rightarrow \infty$).  After finding the
time-evolved state using numerical methods based on \cite{DD03,Hu99} and obtaining the 
reduced density matrices for the two-qubit system by tracing over the environmental variables, 
$\rho_{\rm sys}(t) = {\rm tr}_{\rm env}[\rho(t)]$, they calculated the concurrences for the 
initial Bell states and found that (i) the entanglement evolution was independent of the initial 
state and whether the spin-bath was composed of a finite or infinite number of spins, (ii) a 
smaller anistropy parameter can more easily allow the system to become entangled, (iii) finite 
temperatures allows for ESD, and (iv) large intra-subsystem coupling can help maintain a high 
degree of entanglement.

Ma, Wang, and Cao considered three-qubit systems in a quantum-critical environment consisting of an 
Ising model in a transverse field \cite{MWC07}.  They examined the evolution of the entanglement, 
quantified by the negativity, as a function of system-environment coupling strength, environmental 
degrees of freedom, transverse field strength, and in relation to state symmetry.  The three-qubit states 
considered include: the standard GHZ state $\ket{GHZ} = \frac{1}{\sqrt{2}}(\ket{000} + \ket{111})$, 
the W state   $\ket{W} = \frac{1}{\sqrt{3}}(\ket{001} + \ket{010} + \ket{100})$, and the three-qubit 
Werner state $\rho_{\rm  3-Werner} = \frac{p}{8}\mathbb{I}_{8} + (1-p)\ket{\rm GHZ}\bra{\rm GHZ}$.
For these, they considered the entanglement between for the following bipartitions using the negativity: 
$N_{\rm AB-C}$, $N_{\rm AC-B}$, $N_{\rm BC-A}$, $N_{\rm A-B}$, $N_{\rm A-C}$, $N_{\rm B-C}$, and 
confirmed the existence of ESD relative to these bipartitions as they explored the conditions upon 
which ESD takes place and found that disentanglement can be enhanced by a quantum phase transition 
when the system is coupled weakly to the environment.  Although the lack of entanglement 
in all bipartitions strongly suggests the lack of entanglement at the tripartite level, this is not 
necessarily so; demonstrating this requires a rigorously defined genuinely tripartite mixed state 
entanglement measure, which has yet to be found.
Sun, Wang, and Sun considered a similar model \cite{SWS07}.   
For the two-qubit system with entanglement evolution quantified by the concurrence, 
the initial pure state $\ket{\Phi^+} = \frac{1}{\sqrt{2}}(\ket{00} + \ket{11})$ 
asymptotically disentangles, whereas the initial mixed two-qubit Werner state 
$\rho_{\rm 2-Werner} = p\ket{\Phi^+}\bra{\Phi^+} + \frac{1-p}{4}\mathbb{I}_{4}$ exhibits ESD.
Similarly, the two-qutrit system with entanglement evolution quantified by the negativity, the initial 
pure state $\ket{\Psi} = \frac{1}{\sqrt{3}}(\ket{00} + \ket{11} + \ket{22})$ exhibits asympototic 
disentanglement, either to zero or to some constant, and the initial mixed three-qubit Werner state
$\rho_{\rm 3-Werner} = p\ket{\Psi}\bra{\Psi} + \frac{1 - p}{9}\mathbb{I}_{9}$ becomes completely
disentangled in finite time.  For both the two-qubit and two-qutrit systems, the specific rate of 
asymptotic disentanglement and the actual time of ESD in this situation is determined by several 
factors, including the transverse field strength and the system-environmental coupling factors.  
Nonetheless, the authors demonstrated the basic qualitative result that ESD occurs 
in this model, confirming previous results.

Lai, Hung, Mou, and Chen consider two-qubits coupled to an XXZ spin chain \cite{LHMC08}.  
The novel feature of their analysis was that it goes beyond the usual method involving
a central system coupled to the environment, and instead uses time-dependent density 
renormalization group theory (t-DMRG) to non-perturbatively model spin-bath dynamics 
in order to determine the decoherence and disentanglement properties.  It was shown 
in this way that the phase of the environmental spin bath is important to entanglement 
dynamics.  However, only in the anti-ferromagnetic and paramagentic phases was 
ESD shown to exist.  Using the concurrence, the authors determined the time for ESD for spin-bath 
phase diagram.  Numerous other results were also discovered, demonstrating the relevance 
of quantum phase transitions and ESD.

An important example of the analogous case to spins, that of harmonic oscillator systems, 
is  the a comprehensive analysis of Chou, Yu, and Hu \cite{CYH08}, who demonstrated 
that ESD exists under a wide variety of situations.

%%%%%%%%%%%%%%%%%%%%%%%%%%%%%%%%%%%%%%%%%%%%%%%%%%%%%%%%%%%%%%%%%%%%%%%%%%%%%%%%%%%%%%%%%%%%%%%%%%%
%%%%%%%%%%%%%%%%%%%%%%%%%%%%%%%%%%%%%%%%%%%%%%%%%%%%%%%%%%%%%%%%%%%%%%%%%%%%%%%%%%%%%%%%%%%%%%%%%%%
\subsubsection{Other realizations of ESD}\label{Breadth of ESD}

The study of ESD in abstract models was further developed in more physically realistic contexts
with an eye towards experimental observation and practical application. An example from each
of these contexts is provided in this section.

Cavity quantum electrodynamics (QED), which has been well-developed theoretically and experimentally, 
is well-suited to investigate ESD.  One particular cavity QED model is frequently used in ESD 
studies - that involving two atoms in independent cavities evolving in time under a double Jaynes-Cummings Hamiltonian,
\begin{equation}
H_{\rm JC} = 
\frac{\omega_{0}}{2}\sigma_{z}^{\rm A} + g(a^{\dagger}\sigma_{-}^{\rm A} + \sigma_{+}^{\rm A}a) + \omega a^{\dagger}a +
\frac{\omega_{0}}{2}\sigma_{z}^{\rm B} + g(b^{\dagger}\sigma_{-}^{\rm B} + \sigma_{+}^{\rm B}b) + \omega b^{\dagger}b \ ,
\end{equation}
with $\hbar = 1$, $\omega$ the cavity frequency, $\omega_{0}$ the atomic frequency, and $g$ 
the atom-cavity coupling strength.  This basic Hamiltonian can be modified in a variety of ways, 
such as changing the atom-cavity coupling strengths, specifying various combinations of atomic 
and cavity frequencies, and considering different initial entangled states.  Using this Hamiltonian, 
Y$\ddot{\rm o}$na\c{c}, Yu, and Eberly \cite{YYE07} considered a four-qubit model with a cavity 
QED interpretation, such that all reduced density matrices are of the X-state form, as in Eq. 
\ref{rhoX} with $z = 0$.  Six concurrences were calculated 
($C_{\rm AB}$, $C_{\rm ab}$, $C_{\rm Aa}$, $C_{\rm Bb}$, $C_{\rm Ab}$, and $C_{\rm Ba}$),
allowing them to study ESD and entanglement revivals, with the stipulation that there is no contact 
between either of the two atoms or the two cavities.  Entanglement evolution was considered  
a pure exchange and information-based phenomenon independent of actual physical interaction.

Quantum optics also provides many opportunities in which to better understand ESD.  The study of 
coherence and entanglement in this field has been especially well researched, due to their relation 
to practical quantum communications technologies.  For example,  Gong, Zhang, Dong, Niu, Huang, and 
Guo considered states of single- and double-photon polarization states subject to local phase noise 
\cite{GZD08}, in the form of Eq. \ref{rhoX}, again with $z = 0$.  The decoherence and disentanglement properties 
were examined as a function of different frequency spectrum envelopes.  They discovered that some 
colored frequency spectra lead to asymptotic decoherence and disentanglement and that others can lead to 
periodic sudden deaths and births of coherence and entanglement due to non-Markovian effects.
  
%Lastra, Romero, C. E. L$\acute{\rm o}$pez, N. Zagury, and J. C. Retamal studied entangled coherent 
%states subject to dissipation and mapped onto an orthogonal, time-dependent, finite-dimensional basis 
%\cite{LRL08}.  Using the concurrence, ESD was shown to exist alongside asympototic disentanglement 
%for a variety of initial state amplitudes and relative phases.
%For initial states given by the following form,
%\begin{equation}
%\ket{\Psi} = c_{1}\ket{\alpha}\ket{\beta} + c_{2}\ket{\delta}\ket{\gamma} + c_{3}\ket{\delta}\ket{\beta} + %c_{4}\ket{\alpha}\ket{\gamma} \ ,
%\end{equation}
%where $\ket{\alpha}(\ket{\beta})$ have the same phase as $\ket{\delta}(\ket{\gamma})$ 
%and $c_{i} \in \mathbb{C}$ ($i=1,2,3,4$).

In addition to the discovery and analysis of ESD in cavity QED and quantum optical setups, 
ESD has also been determined to exist in other physical systems.  For example, 
Abdel-Aty showed the existence of ESD in a Josephson junction by modeling a two-qubit maximally 
entangled mixed state, with each qubit of a Cooper pair box connected to a reservoir 
through a Josephson junction \cite{Abdel-Aty08}.  Roszak, Machnikowski, and Jacak showed that 
ESD exists on in a system composed of excitons on quantum dots subject to phonon dephasing noise 
and depends on the distance between these two excitons due to the finite phonon phase 
velocity \cite{RMJ06}.

%%%%%%%%%%%%%%%%%%%%%%%%%%%%%%%%%%%%%%%%%%%%%%%%%%%%%%%%%%%%%%%%%%%%%%%%%%%%%%%%%%%%%%%%%%%%%%%%%%%
%%%%%%%%%%%%%%%%%%%%%%%%%%%%%%%%%%%%%%%%%%%%%%%%%%%%%%%%%%%%%%%%%%%%%%%%%%%%%%%%%%%%%%%%%%%%%%%%%%%
\subsection{ESD: Empirical evidence and experimental proposals}\label{ESD: Empirical evidence and experimental proposals}

Given that entanglement sudden death has been shown to exist in the theoretical context, both in abstract
treatments and in realistic models of specific physical systems, it is not surprising that specific experimental
tests have been proposed and that some of the above results have already been experimentally confirmed.
The experimental observation of entanglement sudden death in physical systems necessarily requires
measurement, often requiring resource-intensive full tomographic state reconstruction.  This is so 
because the construction of entanglement measures may involve operations that do not necessarily 
have one-to-one correspondence to the physical world, for example, the cutoff function used to 
compute the concurrence and partial transpose operation used in the negativity are abstract 
mathematical operations. 

Theoretical support for the measurement, characterization, and 
dynamically monitoring of entanglement evolution has been provided, for example, Santos, 
Milman, Davidovich, and Zagury 
proposed the use of entanglement witnesses to signal the existence or absence of 
entanglement by monitoring a single measurable quantity that remains constant throughout the 
evolution of the quantum system \cite{SMDZ06}.  Furthermore, they proposed an experimental setup that may be 
used for cavity quantum electrodynamics and for trapped ions.  Carvalho, Busse, Brodier, Viviescas, 
and Buchleitner \cite{CBB07} generalized the idea of entanglement measurement and showed that for 
experimentally relevant mixed quantum systems evolving under open quantum systems, entanglement 
monitoring may proceed in an ``optimal'' way based on quantum jump operators.  

Experimental evidence for ESD in a variety of physical contexts such optical setups and atomic ensembles 
has already been given.  Almeida, de Melo, Hor-Myell, Salles, Walborn, Souto Ribeiro, and L. Davidovich 
experimentally confirmed the existence of entanglement sudden death for a two-qubit system due to multi-local 
dephasing and amplitude damping noise in an optical experimental setup involving a Sagnac-like interferometer 
\cite{ADH07}.  In the two-qubit system, one qubit is denoted by the horizontal and vertical polarizations of 
a photon, the other qubit is the ground and excited state of an atom, and the environment acting upon these 
two-qubit system is the momentum of the photon. The general photon polarization Bell-state
$\ket{\Phi} = |\alpha | \ket{HH} + |\beta | \exp(i \delta )\ket{VV}$, where $H$ is horizontal polarization 
and $V$ is the vertical, is considered and may be viewed as a subclass of the X-states, with $w = z = 0$ in 
Eq. \ref{rhoX}.  They consider two initial states: $\ket{\psi_{I}}$  defined by $|\beta |^{2} = |\alpha |^{2}/3$ 
and $\ket{\psi_{II}}$ defined by $|\beta |^{2} = 3 |\alpha |^{2}$.  These initial states both have a concurrence 
of $C(\rho) \approx 0.8$ and similar purity, respectively, 
$\mathcal{P}_{I} \approx 0.91$ and $\mathcal{P}_{II} \approx 0.97$.  $|\alpha |$, $|\beta |$, and $\delta$ are 
modified physically by a combination of quarter- and half-wave plates put in the pump beam path.  The evolution 
of the system for an amplitude damping channel is given by the following map,
\begin{eqnarray}
\ket{H} \otimes \ket{b} & \rightarrow & \ket{H} \otimes \ket{b} \\
\ket{V} \otimes \ket{a} & \rightarrow &
\sqrt{1-p} \ket{V} \otimes \ket{a} + \sqrt{p}\ket{H} \otimes \ket{b} \ ,
\end{eqnarray}
where $\ket{a}(\ket{b})$ denote orthogonal spatial modes.  Entanglement is measured by the
concurrence, given by
\begin{eqnarray}
C(\rho) = \max \left\{ 0, 2(1-p)| \beta | (| \alpha | - p | \beta |) \right\} \ .
\end{eqnarray}
Here, for the case of $| \beta | \leq | \alpha |$, there is no entanglement if $p = 1$.  In contrast, 
for $| \beta | > | \alpha |$, finite-time disentanglement occurs for $p = | \alpha / \beta |$.  They 
showed that $\ket{\psi_{I}}$ undergoes asymptotic disentanglement, with complete disentanglement 
occuring only when $p = 1$, the case where each individual subsystem is completely decohered.  
For $\ket{\psi_{II}}$, the concurrence goes to zero for $p < 1$, showing ESD.  The dephasing channel 
is described by the following map,
\begin{eqnarray}
\ket{H} \otimes \ket{b} & \rightarrow & \ket{H} \otimes \ket{b} \\
\ket{V} \otimes \ket{a} & \rightarrow &
\sqrt{1-p} \ket{V} \otimes \ket{a} + \sqrt{p} \ket{V} \otimes \ket{b} \ .
\end{eqnarray}
Here, both states $\ket{\psi_{I}}$ and $\ket{\psi_{II}}$ exhibit the same behavior; 
only for $p = 1$ do they completely disentangle.  This study was the first
experimental confirmation of ESD.
Salles, de Melo, Almeida, Hor-Meyll, Walborn, Souto Ribeiro, and Davidovic later expanded 
on this discovery by a comprehensive analysis of quantum optical experiments exploring 
further ESD for the amplitude damping channel \cite{SDA08}.

Laurat, Choi, Deng, Chou, and Kimble studied the heralded entanglement between two qubits,
each represented by collective excitations of a cloud of cesium atoms \cite{LCD07}.  For 
purposes of determining the entanglement content, the states of the two atomic ensembles 
are each locally mapped to photonic states.  Since entanglement is a non-increasing quantity 
under local operations, this mapping process ensures that the entanglement in the photonic
states cannot be more than that in the atomic ensembles.  The relevant density matrix is 
of the X-state form of Eq. \ref{rhoX} with $w = 0$, given in the photon number 
($\left\{ n, m \right\} = \left\{ 0, 1\right\}$) basis, $\ket{n}\ket{m}$,
\begin{eqnarray}
\rho = \frac{1}{P}
\left(
\begin{array}{cccc}
 p_{00} & 0 		   & 0      & 0 \\
 0      & p_{01}   & z      & 0 \\
 0      & z^{\ast} & p_{10} & 0 \\
 0      & 0        & 0      & p_{11} 
\end{array}
\right) , \ 
\end{eqnarray}
with $P = p_{00} + p_{01} + p_{10} + p_{11}$ the normalization factor, $p_{ij}$ is the 
probability of $i$ and $j$ ($\left\{ n, m \right\} = \left\{ 0, 1\right\}$) photons 
existing in each of the photonic modes.  The concurrence is then given by
\begin{eqnarray}
C(\rho) = \max[0, C_{0}] \ \ \ \ \ {\rm with} \ \ \ \ \ C_{0} = \frac{1}{P}(2|z| - 2 \sqrt{p_{00}p_{11}} ) \ ,
\end{eqnarray}
Concurrence as a function of storage time begins above zero and goes abruptly to zero in finite time, 
proving the existence of ESD in a system composed of entangled atoms ensembles.

%%%%%%%%%%%%%%%%%%%%%%%%%%%%%%%%%%%%%%%%%%%%%%%%%%%%%%%%%%%%%%%%%%%%%%%%%%%%%%%%%%%%%%%%%%%%%%%%%%%
%%%%%%%%%%%%%%%%%%%%%%%%%%%%%%%%%%%%%%%%%%%%%%%%%%%%%%%%%%%%%%%%%%%%%%%%%%%%%%%%%%%%%%%%%%%%%%%%%%%
%%%%%%%%%%%%%%%%%%%%%%%%%%%%%%%%%%%%%%%%%%%%%%%%%%%%%%%%%%%%%%%%%%%%%%%%%%%%%%%%%%%%%%%%%%%%%%%%%%%
\section{Non-locality Sudden Death}\label{non-locality Sudden Death}

A significant obstacle for the investigation of Entanglement Sudden Death (ESD) for mixed multipartite systems 
has been the lack of a generalized entanglement measure.  However, the investigation of the sudden death of the 
somewhat similar global state property of Bell non-locality has been feasible. In particular, the
finite-time  loss of Bell-non-locality under noise-induced asymptotic state decoherence,  Bell non-locality 
Sudden Death (BNSD), has been theoretically demonstrated, extending research on the destruction of global 
state properties by local noise to three-qubit and larger systems.  This avenue of research may 
also help illuminate the relationship between entanglement and non-locality, which have 
differences the origin of which are not well-understood, even for a two-qubit system.  For example, Werner showed 
that a class of mixed bipartite two-qubit states can be entangled despite not violating a Bell 
inequality \cite{Werner89}, that is, some entangled states may admit a local hidden variables model 
\cite{TA06}. Similarly, it is conceivable that the study of ESD and BNSD could provide a
better understanding of other fundamental questions, such as the reason for the specific value of the Tsirelson 
bound \cite{Tsirelson80,Tsirelson87}.

%%%%%%%%%%%%%%%%%%%%%%%%%%%%%%%%%%%%%%%%%%%%%%%%%%%%%%%%%%%%%%%%%%%%%%%%%%%%%%%%%%%%%%%%%%%%%%%%%%%	
\subsection{Non-locality measures}\label{Non-locality measures}

Bell Non-locality Sudden Death (BNSD) has so far been studied in two- and three-qubit systems.  
For the bipartite two-level system, there exists a single necessary and sufficient condition 
for Bell non-locality, allowing it to be readily studied.  Specifically, non-locality is 
detected by violation of the Clauser-Horne-Shimony-Holt inequality \cite{CHSH69}:
\begin{equation}
\mathcal{B}_{2} = \frac{1}{2}
\left[M_{\rm A}M_{\rm B} + M_{\rm A}M_{\rm B}' + M_{\rm A}'M_{\rm B} - M_{\rm A}'M_{\rm B}'\right] \ ,
\end{equation}
where the operators $M$ and $M'$ act in different directions, $|\mathcal{B}_{2}| \leq 2$ indicates the
existence of a local classical model and $2 < |\mathcal{B}_{2}| \leq 2\sqrt{2}$ denotes the presence 
of quantum correlations.  The upper bound of $2\sqrt{2}$, the Tsirelson bound, specifies the maximum reached by 
quantum mechanics; the algebraic maximum at $|\mathcal{B}_{2}| = 4$ denotes maximal non-locality, 
with the region $2\sqrt{2} < |\mathcal{B}_{2}| \leq 4$ containing non-physical ``super-quantum'' 
correlations that are implied when only a subset of known physical principles are enforced.

For a three-qubit system, both the classification of initially entangled states and the set of 
Bell inequalities used to detect their non-locality become more complex.  (The classification 
methodology and tripartite SLOCC classes are introduced in Sec. \ref{Entanglement classes}.)  
For such systems, tripartite Bell inequalities fall into two categories \cite{Cereceda02}.  
The first category, the Werner-Wolf-$\dot{\rm Z}$ukowski-Brukner (WWZB) inequalities 
\cite{WW01,ZB02}, distinguishes between tripartite states that are describable by a local 
classical model and those that exhibit ``any'' non-locality at all.  The second category, the 
Svetlichny inequalities, distinguish states that can be described by a hybrid local-non-local 
model and those that cannot.  Let us discuss these in turn.

For tripartite states to be describable by a local classical model, the ``entire'' set of 
Werner-Wolf-$\dot{\rm Z}$ukowski-Brukner inequalities must be satisfied \cite{WW01,ZB02}.  
For the tripartite case, there exist five classes of inequalities encompassing thirty-two 
total inequalities, with the inequalities contained within each class equivalent upon 
symmetries.  Because the conjunctions of these inequalities are a necessary and sufficient 
condition for non-locality, the satisfaction of ``all'' these inequalities is necessary and 
sufficient to demonstrate a local classical model, however, only a violation of ``one'' is 
sufficient to show the existence of non-locality.  They are:
\begin{eqnarray}
{\rm (P1)}\ \ \ \ \ \ \mathcal{B}_{{\rm P}1}&=& 
2 M_{\rm A}M_{\rm B}M_{\rm C} \ , \nonumber \\
{\rm (P2)}\ \ \ \ \ \ \mathcal{B}_{{\rm P}2}&=& 
\frac{1}{2}(- M_{\rm A}M_{\rm B}M_{\rm C}   + M_{\rm A}M_{\rm B}M_{\rm C}'
+ M_{\rm A}M_{\rm B}'M_{\rm C}  + M_{\rm A}M_{\rm B}'M_{\rm C}' \nonumber \\
&+& M_{\rm A}'M_{\rm B}M_{\rm C}  + M_{\rm A}'M_{\rm B}M_{\rm C}' 
+ M_{\rm A}'M_{\rm B}'M_{\rm C} + M_{\rm A}'M_{\rm B}'M_{\rm C}'
)\ , \nonumber \\
{\rm (P3)}\ \ \ \ \ \ \mathcal{B}_{{\rm P}3}&=&
[M_{\rm A}(M_{\rm B} + M_{\rm B}') + M_{\rm A}'(M_{\rm B} - M_{\rm B}') ]M_{\rm C}\ , \nonumber \\
{\rm (P4)}\ \ \ \ \ \ \mathcal{B}_{{\rm P}4}&=& 
M_{\rm A}M_{\rm B}(M_{\rm C} + M_{\rm C}') - M_{\rm A}'M_{\rm B}'(M_{\rm C} - M_{\rm C}')\ , \nonumber \\
{\rm (P5)}\ \ \ \ \ \ \mathcal{B}_{{\rm P}5}&=& 
M_{\rm A}M_{\rm B}M_{\rm C}' + M_{\rm A}M_{\rm B}'M_{\rm C} +
M_{\rm A}'M_{\rm B}M_{\rm C} - M_{\rm A}'M_{\rm B}'M_{\rm C}' \ ,
\end{eqnarray}
where $M_{i}$ ($i= A,B,C$) are projective measurement operators acting on the $i^{\rm th}$ 
qubit and the prime denotes an alternate measurement.  All of these must be satisfied for 
a state to be describable by a local classical model,
$|\left\langle \mathcal{B}_{{\rm P} I} \right\rangle_{\rho}| \leq 2$ ($I$ = 1, 2, 3, 4, 5).  
This set includes the often considered sub-class of tripartite Bell inequalities 
\cite{Mermin90,Ardehali92,BK93}, namely those involving $\mathcal{B}_{{\rm P}5}$.

Although the WWZB set of Bell inequalities provide necessary and sufficient conditions 
for determining whether a state can be described by a local classical model, they cannot 
distinguish between whether a state is describable by a truly non-local model or a hybrid 
local-non-local model.  For this, one needs to consider the class of Svetlichny inequalities, 
represented by
\begin{eqnarray}
{\bf S} &=& 
M_{\rm A}M_{\rm B}M_{\rm C}  + M_{\rm A}M_{\rm B}M_{\rm C}' + 
M_{\rm A}M_{\rm B}'M_{\rm C} + M_{\rm A}'M_{\rm B}M_{\rm C} \nonumber \\
&-& M_{\rm A}'M_{\rm B}'M_{\rm C}' - M_{\rm A}'M_{\rm B}'M_{\rm C} - 
M_{\rm A}'M_{\rm B}M_{\rm C}' - M_{\rm A}M_{\rm B}'M_{\rm C}' \ , 
\end{eqnarray}
where the measurement operators $M(M')$ are defined as before, $|{\bf S}| > 4$ denotes 
genuine tripartite Bell non-locality, $|{\bf S}| = 4\sqrt{2}$ is the maximum attainable
through quantum mechanical correlations, and $|{\bf S}| = 8$ is the algebraic maximum 
\cite{Svetlichny87}.  (Note that there exists another Svetlichny operator expectation 
$S'$ that may contain have a different phase, but it is equivalent to this one when 
only the magnitude of the quantity is considered.)
%%%%%%%%%%%%%%%%%%%%%%%%%%%%%%%%%%%%%%%%%%%%%%%%%%%%%%%%%%%%%%%%%%%%%%%%%%%%%%%%%%%%%%%%%%%%%%%%%%%		
\subsection{Specific BNSD studies}\label{Specific BNSD studies}

Ann and Jaeger \cite{AJ08b} were first to consider the finite-time loss of Bell-inequality 
violation (BNSD) in a three-qubit system.  They considered a pure generalized W state,
\begin{equation}
\ket{W_{g}} = \bar{a}_{1}\ket{001} + \bar{a}_{2}\ket{010} + \bar{a}_{4}\ket{100} \ ,
\end{equation}
with coefficients $\bar{a}_{i} \in \mathbb{C}$ ($i=1,2,4$) satisfying $\sum_{i}\bar{a}_{i}^{2} = 1$. 
Using the operator-sum decomposition, the time evolution was considered in an analogous way to 
previous ESD studies, initial state is considered subject to multi-local dephasing noise, which 
affects the state density matrix by the appearance of multiplicative exponential decay factors 
$\gamma(t) = e^{- \tilde{\Gamma} t}$ on the off-diagonal elements, where $\tilde{\Gamma}$ denotes 
the strength of local dephasing.  For an asymptotically decohering state, the finite-time loss 
of Bell-non-locality is demonstrated when ``both'' of the following conditions involving the two 
different (non-)locality regimes are satisfied:
\begin{eqnarray}
{\rm (i)} && \left\langle \mathcal{B}_{2} \right\rangle_{\rho(0)} > 2 \\
{\rm (ii)} && \left\langle \mathcal{B}_{2} \right\rangle_{\rho(t)} \leq 2 \ , 
\end{eqnarray}
with $t_{\rm 2} \leq t < \infty$, $t_{2}$ denoting the time a violation 
ceases to exist.  The specific measurement operators used for qubits A, B, and C were
\begin{eqnarray}
M_{\rm A} &=& \sigma_{z} \otimes \mathbb{I} \otimes \mathbb{I} \ , \\
M_{\rm A}' &=& \sigma_{x} \otimes \mathbb{I} \otimes \mathbb{I} \ , \\
M_{\rm B}  &=& \mathbb{I} \otimes
\left[\cos\left(\theta_{\rm B}\right)\sigma_{z}-\sin\left(\theta_{\rm B}\right)\sigma_{x}\right]
\otimes \mathbb{I} \ , \\
M_{\rm B}' &=& \mathbb{I} \otimes
\left[\sin\left(\theta_{\rm B}\right)\sigma_{z}+\cos\left(\theta_{\rm B}\right)\sigma_{x}\right]
\otimes \mathbb{I} \ , \\
M_{\rm C}  &=& \mathbb{I} \otimes \mathbb{I} \otimes
\left[\cos\left(\theta_{\rm C}\right)\sigma_{z}-\sin\left(\theta_{\rm C}\right)\sigma_{x}\right] \ , \\
M_{\rm C}' &=& \mathbb{I} \otimes \mathbb{I} \otimes
\left[\sin\left(\theta_{\rm C}\right)\sigma_{z}+\cos\left(\theta_{\rm C}\right)\sigma_{x}\right] 
\label{generalTripartiteOperators} \ ,
\end{eqnarray}
where $\sigma_{z}$ and $\sigma_{x}$ are Pauli operators, with the rotational angles given by 
the specific values $\theta_{\rm B} = \frac{\pi}{6}$ and $\theta_{\rm C} = \frac{\pi}{3}$.
For the maximally entangled case where the coefficients of the $\ket{W_{g}}$ are equal and real,
$\bar{a}_{1} = \bar{a}_{2} = \bar{a}_{4} = 1 / \sqrt{3}$, it was shown that complete loss of Bell 
inequality violation occurs at the finite time scale
\begin{equation}
t_{2} = \frac{\ln (2)}{2 \Gamma} \ .
\end{equation}

\begin{figure}[htbp]
	\centering
	\includegraphics[width=0.6\textwidth]{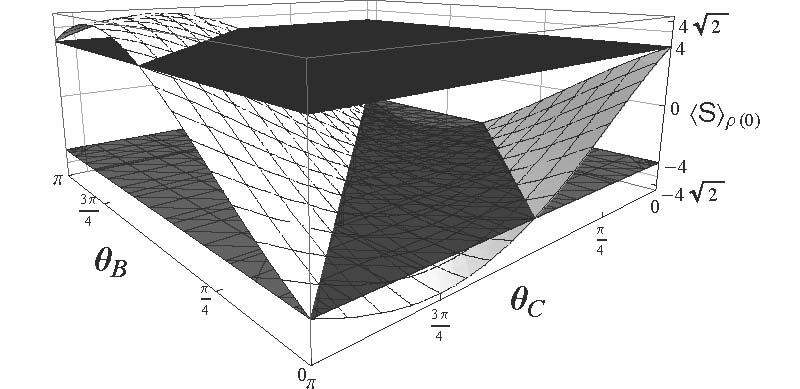}
	\caption{At initial time, the Svetlichny expectation $\left\langle{\bf S}\right\rangle_{\rho(0)}$ 
	is plotted as a function of the rotation angles $\theta_{\rm B}$ and $\theta_{\rm C}$ for the 
	maximally entangled tripartite state, $\ket{GHZ_{3}} = (\ket{000} + \ket{111})/\sqrt{2}$.  
	$|\left\langle{\bf S}\right\rangle_{\rho(0)}| >  4$ denotes violation of the Svetlichny inequality 
	and genuinely tripartite Bell non-locality.
	}
	\label{Svetlichny1}
\end{figure}

\begin{figure}[h]
	\centering
		\includegraphics[width=0.40\textwidth]{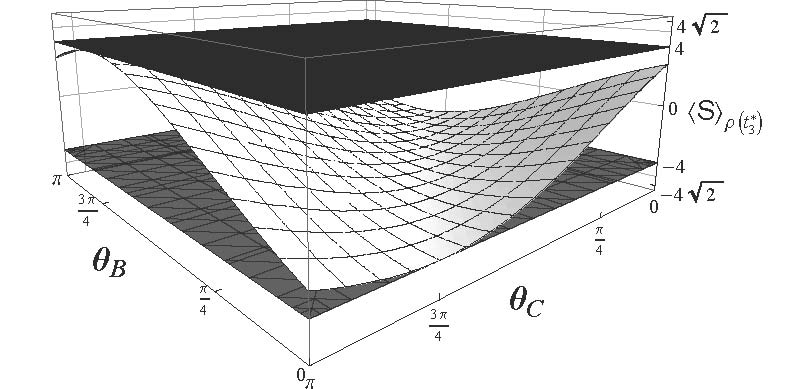}
		\includegraphics[width=0.40\textwidth]{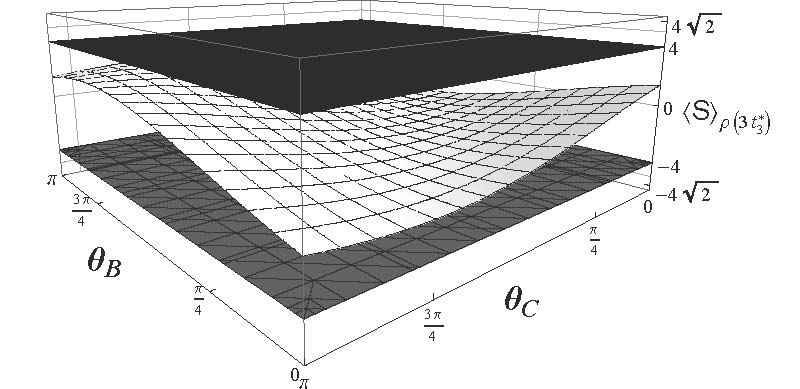}
	\caption{
	These plots show subsequent time evolution of the Svetlichny expectation as the state is subject 
	to multi-local dephasing noise.  On the left, $|\left\langle{\bf S}\right\rangle_{\rho(t_{3}^{*})}| \leq 4$ 
	for all measurement angles $\theta_{\rm B}$ and $\theta_{\rm C}$, indicating the lack of genuinely 
	tripartite Bell nonlocality.  Because this occurs in finite time $t_{3}^{*}$, despite violation at 
	$t=0$ and asymptotic decoherence, genuine tripartite Bell non-locality sudden death is shown to exist.
	Subsequent time evolution is shown on the right, for example at $t = 3t_{3}^{*} > t_{3}^{*}$, 
	$|\left\langle{\bf S}\right\rangle_{\rho(t)}| \leq 4$.	The set of WWZB inequalities exhibit similar 
	behavior, denoting the loss of ``all'' generic Bell non-locality in finite time.
	}
	\label{Svetlichny2and3}
\end{figure}

In a follow-up publication \cite{AJ08c}, the authors of \cite{AJ08b} improved upon the above initial 
study of the finite-time loss of Bell-inequality violation in W-states by considering the effects of 
multi-local dephasing noise acting on the initial generic class of tripartite entangled states, 
\begin{equation}
\ket{\Psi_{3}} = \bar{a}_{0}\ket{000} + \bar{a}_{4}\ket{100} + \bar{a}_{5}\ket{101} + 
\bar{a}_{6}\ket{110} + \bar{a}_{7}\ket{111} \ , 
\end{equation}
with a more nuanced examination of non-locality properties using the entire set of WWZB and 
Svetlichny inequalities.
Genuinely tripartite non-locality was considered first, using the Svetlichny inequality.  Here, 
the maximum quantum violation  $| \left\langle {\bf S} \right\rangle | = 4\sqrt{2}$, is reached 
by the GHZ state with $|\bar{a}_{0}| =  |\bar{a}_{7}| = 1/\sqrt{2}$ and all other coefficients 
zero.  Under multi-local dephasing noise and for the set of measurement operators in Eq.
\ref{generalTripartiteOperators} chosen for maximum initial time violation of $S$, the 
time-dependent Svetlichny expectation of the three-qubit state is given by
\begin{equation}
|\left\langle S \right\rangle_{\rho(t)}| = 8\sqrt{2}|\bar{a}_{0}||\bar{a}_{7}|e^{-3\Gamma t} \ , 
\end{equation}
where $\Gamma$ parameterizes the strength of the dephasing noise.  Tripartite non-locality for 
the maximally entangled GHZ state with $\bar{a}_{0} = \bar{a}_{7} = 1/\sqrt{2}$ and all other 
coefficients zero, as distinguished by the Svetlichny inequality, is completely lost at a time
\begin{equation}
t_{3}^{\ast} = \frac{\ln (\sqrt{2})}{3\Gamma} \ ,
\end{equation}
and for all times afterwards.  
Figs. \ref{Svetlichny1} and \ref{Svetlichny2and3} provide a visualization this time evolution.

%%%%%%%%%%%%%%%%%%%%%%%%%%%%%%%%%%%%%%%%%%%%%%%%%%%%%%%%%%%%%%%%%%%%%%%%%%%%%%%%%%%%%%%%%%%%%%%%%%%%

Next, the loss of ``all'' non-locality was considered using the set of WWZB inequalities.
The maximum quantum violation possible for the set of WWZB inequalities is
$|\left\langle \mathcal{B} \right\rangle| = 4$.  Determining the non-locality properties 
of using the WWZB inequalities requires examining each of the five independent classes 
of inequalities, however, the time required for complete loss of Bell-inequality violation 
was found to be bounded from above by the time it takes $\mathcal{B}_{{\rm P}5}$ to be completely 
lost.  Therefore, the relevant quantity that captures Bell non-locality decay of the W 
state is given by
\begin{equation}
| \left\langle B_{{\rm P}5} \right\rangle_{\rho(t)}| = 8|\bar{a}_{0}||\bar{a}_{7}|e^{-3\Gamma t} \ .
\end{equation}
The ``entire'' set of WWZB inequalities was shown to be completely lost in the finite time scale
\begin{equation}
t_{3} = \frac{\ln(2)}{3\Gamma} \ ,
\end{equation}
and for all times afterwards.  Thus, the tripartite state was shown to be described 
by a local classical model.  This analysis used the Svetlichny inequality 
to demonstrate the loss of genuinely tripartite non-locality and used the entire set of 
WWZB inequalities to show the transition from a non-local state to a local classical state;  
despite state decoherence occuring asymptotically in time, Bell non-locality was completely 
lost in finite time.

Tripartite BNSD has very recently been applied in other contexts, making use of a similar 
methodology as the initial discovery.  Yang, Yang, and Cao 
considered a system composed of two identical atoms in a lossy cavity according to the 
Tavis-Cummings model \cite{YYC08}.  The initial entangled state for the tripartite system is 
composed of the two atoms and the cavity, was taken to be the W state, 
$\ket{\psi(0)} = a\ket{eg0} + b\ket{ge0} + c\ket{gg1}$, where $\ket{g}$ and $\ket{e}$ denote
the ground and excited states of the atoms, respectively, $\ket{0}(\ket{1})$ denotes the 
two different states of the cavity, and $a^{2} + b^{2} + c^{2} = 1$.  The non-locality properties 
are exmained as a function of time for the maximally entangled W state with $a = b = c = 1/\sqrt{3}$.
They calculated the critical times for loss of Bell inequality violation for different relative 
values of the photon leakage rate $\kappa$ and the coupling constant $\lambda$; 
$t(\kappa = \lambda) > \ln(2)/\kappa$ and $t(\kappa = \lambda) = \ln(2)/\kappa$; when 
$\kappa < \lambda$, the tripartite non-locality oscillates before decays asympototically.  
Note that there can exist non-locality revivals at the boundary.

Qiu, Wang, Su, and Ma similarly considered a tripartite state of spins coupled to an antiferromagnetic 
environment with an applied magnetic field \cite{QWSM08}.  The Hamiltonian is derived using the spin-wave
approximation.  For an initial separable system composed of the tensor product of an environment in 
thermal equilbrium with a pure tripartite GHZ state, these authors found the time until BNSD as a function 
the applied magnetic field.  When the initial system is a W state, there is no BNSD since the state is a
decoherence free state in that environment.  For this reason, they have suggested that the generalized 
W state is a good candidate for quantum information processing purposes.

Although no experimental studies have yet been performed for the specific purpose of testing the existence 
BNSD, since it was first theoretically explored only in 2008, it is likely that some past experiments 
designed to verify the existence of non-locality itself but failed to definitively do so were
experiments in which BNSD took place during between state preparation and state measurement.	

%%%%%%%%%%%%%%%%%%%%%%%%%%%%%%%%%%%%%%%%%%%%%%%%%%%%%%%%%%%%%%%%%%%%%%%%%%%%%%%%%%%%%%%%%%%%%%%%%%%
\section{Implications and summary}\label{Implications and summary}

The above review has considered a wide range of theoretical situations and a number of experimental 
situations in which entanglement sudden death (ESD) and Bell non-locality sudden 
death (BNSD) have been demonstrated.  These phenomena have been explored only for the last
five years in the first instance and one year in the second, leaving considerable room for further study.  
The existing results already suggest that previous 
studies involving asympototic ``decoherence'' are readily extendable to the consideration  
of disentanglement and the loss of non-locality in finite time in a way that may bear on the foundations
in ways that the study of decoherence has so far failed to, despite initially high expectations for that 
earlier related program of study.

\end{document}